\documentclass[fleqn,usenatbib]{mnras}
\usepackage{fix-cm}

\usepackage{newtxtext,newtxmath}

\usepackage[T1]{fontenc}

\DeclareRobustCommand{\VAN}[3]{#2}
\let\VANthebibliography\thebibliography
\def\thebibliography{\DeclareRobustCommand{\VAN}[3]{##3}\VANthebibliography}

\usepackage{graphicx}	
\usepackage{amsmath}	
\usepackage{subcaption}
\usepackage{multicol}
\usepackage{hyperref}
\usepackage{scalerel}
\usepackage{tikz}
\usepackage{subfiles}
\usepackage{booktabs}
\usepackage{multirow}
\usetikzlibrary{svg.path}
\usepackage{siunitx}
\definecolor{orcidlogocol}{HTML}{A6CE39}
\tikzset{
  orcidlogo/.pic={
    \fill[orcidlogocol] svg{M256,128c0,70.7-57.3,128-128,128C57.3,256,0,198.7,0,128C0,57.3,57.3,0,128,0C198.7,0,256,57.3,256,128z};
    \fill[white] svg{M86.3,186.2H70.9V79.1h15.4v48.4V186.2z}
                 svg{M108.9,79.1h41.6c39.6,0,57,28.3,57,53.6c0,27.5-21.5,53.6-56.8,53.6h-41.8V79.1z M124.3,172.4h24.5c34.9,0,42.9-26.5,42.9-39.7c0-21.5-13.7-39.7-43.7-39.7h-23.7V172.4z}
                 svg{M88.7,56.8c0,5.5-4.5,10.1-10.1,10.1c-5.6,0-10.1-4.6-10.1-10.1c0-5.6,4.5-10.1,10.1-10.1C84.2,46.7,88.7,51.3,88.7,56.8z};
  }
}

\newcommand\orcidicon[1]{\href{https://orcid.org/#1}{\mbox{\scalerel*{
\begin{tikzpicture}[yscale=-1,transform shape]
\pic{orcidlogo};
\end{tikzpicture}
}{|}}}}

\title[Stellar outflows in simulated galaxies]{Emission line tracers of galactic outflows driven by stellar feedback in simulations of isolated disk galaxies}

\author[E. L. Howatson et al.]{
Elliot L. Howatson$^{\orcidicon{0009-0005-8190-6529}}$$^{1,2}$\thanks{E-mail: elliot.howaton-2022@hull.ac.uk},
Alexander J. Richings$^{\orcidicon{0000-0003-0502-9235}}$$^{1,2}$,
Elke Roediger$^{\orcidicon{0000-0003-2076-6065}}$$^{1}$,
Claude-André Faucher-Giguère$^{\orcidicon{0000-0002-4900-6628}}$$^{3}$,
\newauthor Tom Theuns$^{\orcidicon{0000-0002-3790-9520}}$$^{4}$, 
Yuankang Liu$^{\orcidicon{0000-0003-0771-763X}}$$^{4}$, 
Tsang Keung Chan$^{\orcidicon{0000-0003-2544-054X}}$$^{5}$,
Oliver Thompson$^{\orcidicon{0009-0008-9039-9970}}$$^{1, 2}$,
Cody Carr$^{\orcidicon{0000-0003-4166-2855}}$$^{6,7}$ and
\newauthor Daniel Anglés-Alcázar$^{\orcidicon{0000-0001-5769-4945}}$$^{8,9}$
\\
$^{1}$E.A Milne Centre for Astrophysics, University of Hull, Cottingham Road,
 Hull HU6 7RX, UK\\
$^{2}$Centre for Data Science, Artificial Intelligence and Modelling, University of Hull, Cottingham Road, Hull, HU6 7RX, UK\\
$^{3}$ CIERA and Department of Physics and Astronomy, Northwestern University, Evanston, Illinois, USA \\
$^{4}$ Institute for Computational Cosmology, Department of Physics, Durham University, South Road, Durham DH1 3LE, UK \\
$^{5}$ Department of Physics, The Chinese University of Hong Kong, Shatin, Hong Kong, China \\
$^{6}$ Center for Cosmology and Computational Astrophysics, Institute for Advanced Study in Physics, Zhejiang University, Hangzhou 310058,  China \\
$^{7}$ Institute of Astronomy, School of Physics, Zhejiang University, Hangzhou 310058,  China \\
$^{8}$ Department of Physics, University of Connecticut, 196 Auditorium Road, U-3046, Storrs, CT 06269-3046, USA \\
$^{9}$ Center for Computational Astrophysics, Flatiron Institute, 162 5th Avenue, New York, NY 10027, USA
\date{Accepted XXX. Received YYY; in original form ZZZ}
}
\pubyear{2024}

\begin{document}
\label{firstpage}
\pagerange{\pageref{firstpage}--\pageref{lastpage}}
\maketitle

\begin{abstract}
Galactic outflows strongly influence galactic evolution and have been detected in a range of observations. Hydrodynamic simulations can help interpret these by connecting direct observables to the physical conditions of the outflowing gas. Here we use simulations of isolated disk galaxies ranging from dwarf mass ($M_{200} = 10^{10}\mathrm{M}_{\odot}$) to Milky Way mass ($M_{200} = 10^{12}\mathrm{M}_{\odot}$), based on the FIRE-2 subgrid models to investigate multiphase galactic outflows. We use the CHIMES non-equilibrium chemistry module to create synthetic spectra of common outflow tracers ([C\textsc{ii}]$_{158\rm{\mu m}}$, $\mathrm{CO}_{J(1-0)}$, H$\alpha$ and $[\mathrm{O}\textsc{iii}]_{5007\text{\AA}}$). Using our synthetic spectra we measure the mass outflow rate, kinetic power and momentum flux using observational techniques. In [C\textsc{ii}]$_{158\rm{\mu m}}$ we measure outflow rates of $10^{-4}$ to $1$ $\mathrm{M_{\odot}yr^{-1}}$ across an SFR range of $10^{-3}$ to $1$ $\text{M}_{\odot}\text{yr}^{-1}$, which is in reasonable agreement with observations. The significant discrepancy is in $\mathrm{CO}_{J(1-0)}$, with the simulations lying $\approx1$ dex below the observational sample. We test observational assumptions used to derive outflow properties from synthetic spectra. We find the greatest uncertainty lies in measurements of electron density, as estimates using the SII doublet can overestimate the actual electron density by up to 2 dex, which changes mass outflow rates by up to 4 dex. We also find that molecular outflows are especially sensitive to the conversion factor between CO luminosity and H2 mass, with outflow rates changing by up to 4 dex in our least massive galaxy. Comparing the outflow properties derived from the synthetic spectra to those derived directly from the simulation, we find that [C\textsc{ii}]$_{158\rm{\mu m}}$ probes outflows at greater distances from the disk, whilst we find that molecular gas does not survive at large distances within outflows within our modestly star-forming disk galaxies simulated in this work. 
\end{abstract}



\section{Introduction}
\label{intro}
Galactic outflows are a crucial component in the formation and evolution of galaxies and the gas within them, the interstellar medium (ISM) \citep{c+c1985, veilleux2005}. By removing available gas and heating what remains, a galaxy's star formation is regulated or quenched by depleting its star-forming fuel \citep{murray2005, bower2006, bower2017, mitchell2020, muratov2015, pandya2021}. The metal rich nature of galactic outflows allows them to enrich the interstellar medium (ISM) and the circum-galactic medium (CGM), whilst determining the chemical evolution of their host galaxy depending on the amount of outflowing material retained within the galaxy's potential \citep{turner2015, turner2016, muratov2017}. This enrichment allows galactic outflows to shape the mass-metallicity relation within galaxies \citep{ma2016, bassini2024, mars2024}. The tight relationship between mass and metallicity is thought to arise from enriched gas being removed from galaxies via outflows which can more easily escape lower mass halos \citep{tremonti2004, ma2016, chisholm2018}. Gas which makes up what we define as outflows may not necessarily escape the galaxy's halo entirely, and may fall back inwards, creating a galactic "fountain". As material flows back into the galaxy it is recycled into further star formation, and therefore galactic outflows play a key role as the engine which drives the baryon cycle \citep{angelalc2017}.

Outflow properties can be measured observationally either through emission or absorption lines. From these lines we can derive properties such as outflow velocity, temperature, metallicity and mass. The outflowing gas consists of multiple phases with different densities and temperatures \citep{vijayan2024} and, as a result, multiple tracers are needed to build up a full physical picture of the gas in a way that quantifies important properties such as energy, mass and momentum. 

An outflow's cold gas phase (<$10^{4}$K) can be traced by both neutral atomic gas and cool molecular tracers. The neutral atomic tracers include C\textsc{ii}, HI, OI and MgI among others, and the molecular tracers include CO, OH, HCO+ (see \citet{veilleux2020} and references therein). Warm ionised phases at temperatures $\approx10^{4}$K, can be traced by $[\mathrm{O}\textsc{iii}]_{5007\text{\AA}}$ and H$\alpha$ emission. The hottest phases ($>10^{7}$K) give off X-ray emission and can be traced by, for example, iron \citep{turner2008}.

The processes which drive the outflowing gas can be split into two main channels: stellar feedback, which is this work's focus, and active galactic nuclei (AGN). Stellar feedback involves the injection of energy and momentum into the surrounding ISM from a variety of stellar processes, namely supernovae (see \citet{veilleux2005} for a review), stellar winds, radiative pressure and cosmic rays \citep{mcleod2021, krumholz2014, girichidis2020, hopkinsCR2021}. Supernovae inject large amounts of energy into the ISM and is the stellar mechanism most associated with driving outflows from a galaxy. The alternative driving mechanisms inhibit star formation by creating turbulence in the ISM, and heating the gas to prevent further star formation, and radiative feedback especially can modulate mass outflows by reducing SFR in dwarf galaxies \citep{rey2025}. At lower halo masses ($M_{200} \leq 10^{12}\text{M}_{\odot}$) stellar feedback is the primary set of mechanisms despite driving less energetic, lower velocity outflows. This is because at lower halo masses these slower outflows can still escape the galaxy's gravitational potential and star formation activity inhibits accretion onto the central black hole, which in turn prevents the black hole from growing preventing AGN feedback \citep{AA2017}. 

Understanding the physics behind galactic outflows is crucial to explaining the processes driving galaxy evolution, and having a complete set of multiphase observations of outflows can help us achieve this goal. M82 provides the classic example of a galaxy for which a complete set of observations is available \citep{strickland1997, westmoquete2009, cazzoli2016, ott2004, walter2002}. Recent studies are also building up extensive samples of multiphase measurements of outflows in other galaxies (for example \citealt{fluetsch2021}). To interpret these observations we need to develop a theoretical framework of galactic outflows and this begins with analytical models for galactic outflows in \citet{chevalier+clegg1985} before moving onto modern analytical models such as \citet{king2011} and \citet{f-g2012}. We can also use hydrodynamical simulations to complement both analytical models and observations of outflows. By modelling galaxies we can track their evolution from their beginning to end in a way that is simply not feasible for observers. Tracking a galaxy's evolution helps us understand the physics behind observed phenomena which is especially useful as galactic wind observations depend on a number of assumptions such as outflow location, geometry, and host galaxy properties \citep{chisholm2017, RVS2005}, which our simulations can test.

Within the literature there is a broad range of simulations which can be used to study galactic outflows. Starting on the largest scale there are cosmological large-volume simulations, such as EAGLE \citep{joop2015, mitchell2020} and TNG50 \citep{nelson2019}. \citet{mitchell2020} find that the mass loading factor ($\dot{M}_{\mathrm{out}}/SFR$), used to denote outflow strength, scales with the circular velocity as $V_{c}^{2/3}$ for galaxies where stellar feedback dominates ($M_{200}<10^{12}\,M_{\mathrm{\odot}}$). EAGLE does not directly resolve the ISM physics and appears to favour smaller outflows which eventually entrain more gas \citep{mitchell2020}. This is also the case with TNG50; a cosmological, magnetohydrodynamical simulation which is part of the IlllustrisTNG project \citep{springel2018}. Whilst it manages to create stellar driven galaxy outflows, the simulation is very computationally demanding in comparison to others due to its large scale, and the outflows themselves miss the cold gas phases as the implemented equation of state prevents gas cooling below $\sim10^{4}$K. \citet{mitchell2020} also find that low mass galaxy outflows are driven by star formation, and that the velocity of these outflows increase from 200 km/s to 1000 km/s across the total stellar mass range  $M_{*}=10^{7.5}\,\mathrm{M_{\odot}}$ to $M_{*}=10^{11}\,\mathrm{M_{\odot}}$.  Next, there are cosmological zoom in simulations, which look at individual halos. These include, for example, FIRE-2 \citep{hopkins2018a, pandya2021}, AURIGA \citep{grand2016}, APOSTLE \citep{sawala2016} and LYRA \citep{gutcke2021}. \citet{kelly2022} compare the baryon cycles in AURIGA and APOSTLE and the effects of stellar feedback in both of these. They find that the AURIGA simulations are dominated by "ejective" feedback, where star forming gas is removed from the galaxy haloes by strong star-formation driven feedback. APOSTLE on the other hand prevents star formation by driving large outflows which inhibit gas accretion into the galaxy. Finally there are simulations of idealised isolated galaxies and ISM box simulations, for example SILCC \citep{walch2015} and TIGRESS \citep{kim+ostiker2020} which operate on the parsec scale at higher resolutions. This category also includes the QUOKKA simulations \citep{vijayan2024quokka, vijayan2025} which have recently investigated outflows in a simulated patch of gas within a Milky Way mass galaxy.

Our goal in this paper is to use synthetic spectra of observable emission lines from simulated galaxies to measure outflow properties, enabling us to directly compare simulation predictions with observational data. We then use the simulations to test common assumptions used in observational methods deriving outflow properties from spectra. Finally, we calculate outflow properties directly from gas motions within the simulations and compare these to the observationally derived outflow properties. We use simulations of isolated disk galaxies presented in \citet{richings2022} which use the FIRE-2 sub-grid models \citep{hopkins2018a} and the CHIMES non-equilibrium chemistry module for the time-dependent evolution of ions and molecules \citep{richings2014a, richings2014b}.

The paper is structured as follows. Section \ref{method} provides an overview of the simulated galaxies, the non-equilibrium chemistry and how outflow properties can be measured for a variety of emission lines using this chemistry. Section \ref{synthetic sepectra results} presents the results of these emission lines, how they compare to observations of emission lines and the temperature-density phase space properties of each emission line tracer. Section \ref{outflow props from particles} will then calculate outflow properties using phase space cuts, and the particles from the simulations directly before comparing the two methods and their differences before concluding in section \ref{conclusion}. 

\section{Method}
\label{method}
\subsection{Isolated galaxy simulations}
\label{isolated galaxies}
The simulations used here are a set of isolated disk galaxy simulations introduced by \citet{richings2022}. These simulations use the meshless finite mass (MFM) hydro solver implemented in \textsc{gizmo} \citep{hopkins2015} and represent a set of typical star forming galaxies across a range of masses from dwarf galaxies to Milky Way mass galaxies (see section \ref{initial conditions}). The simulations use the sub-grid models for galaxy formation that were developed for the "Feedback in Realistic Environments" (FIRE-2) \footnote{https://fire.northwestern.edu/} simulations \citep{hopkins2018a}. We summarise these models in Section \ref{initial conditions}. \citet{richings2022} coupled these sub-grid models to the \textsc{chimes} non-equilibrium chemistry package to track the evolution of ions and molecules in the isolated galaxy simulations that we use in this work. We summarise methods used for this chemical modelling in Section \ref{chimes}. 

\subsubsection{Initial conditions}
\label{initial conditions}
There are seven simulated galaxies in this work whose halo masses span from dwarf ($M_{200}=10^{10}\text{M}_{\odot}$) to Milky Way mass galaxies ($M_{200}=10^{12}\text{M}_{\odot}$), where $M_{200}$ is the halo mass enclosed within a spherical overdensity at 200 times the critical density. Further properties for these galaxies can be found in table \ref{galaxy props}. The initial conditions of these galaxies were created using \textsc{MAKEDISK} \citep{springel2005}. These galaxies are comprised of three key components: a central stellar bulge, a surrounding disk made up of gas and stars, and a dark matter halo within which the structure is embedded. The bulge and halo follow a \cite{herquist1990} density profile and the stellar/gas disk follow a radial exponential profile. Vertically, the stellar portion of the disk is an isothermal sheet set to 0.1 $\times$ the radial exponential length. The gaseous disk component has its vertical profile computed to be at hydrostatic equilibrium for a given gravitational potential at $10^4$K. Structural parameters for the isolated galaxies were determined based on redshift zero scaling relations for the dark matter halo concentration \citep{duffy2008}, total stellar mass \citep{moster2013, sawala2015}, bulge-to-total stellar mass ratio \citep{benson2007, salo2015} and gas fraction \citep{leroy2008}. 

In our initial conditions the gas begins in the galaxy disc with no gaseous halo. This means that the outflows initially propagate into a vacuum, whilst interacting with previously ejected gas at later time steps. Because of this we differ from cosmological simulations in that we do not model a circum-galactic medium (CGM), which is a limitation of our simulations as the CGM will have an effect on the enrichment and the strength of outflows at halo scales \citep{fielding2017, stern2021}. We therefore focus on outflows as they are being launched, at scales a few kpc from the disc, where interactions with the CGM play less of a role.

The baryonic and dark matter particles have initial masses of $400\,\mathrm{M_{\odot}}$ and $1910\,\mathrm{M_{\odot}}$, respectively. \citet{richings2022} also tested simulation runs at 8x higher and lower particle mass, and they found the luminosities of common ISM emission tracers are not strongly affected by the numerical resolution (see Fig. B4 of \citealt{richings2022}). Therefore, we focus on the fiducial runs in this work.

The initial conditions are first evolved for 300\,Myr with a reduced supernova timescale, which enables the galaxy to settle without disrupting the gaseous disk (see \citet{richings2022} for a full explanation of technical details). The simulations are then run using fiducial models for another 500\,Myr, and it is this period that is used for analysis in our work. The metallicity of the simulated gas and stars is set using the SDSS mass-metallicity relation \citep{andrewsmartini2013}, however the metallicities increase during the initial 300\,Myr evolution so that by the start of the 500\,Myr period the metallicities are 0.004-0.22 dex higher than the observed mass-metallicity relation (see table 1 in \citealt{oliver2024}).

\begin{table*}
    \centering
    \begin{tabular}{cccccc}
         \hline
         Name & Halo Mass $M_{200}$ & Stellar mass $M_{*}$ & Disk Radius $R_{disc}$ & Gas Fraction & B/T\\
         & $\mathrm{M_{\odot}}$ & $\mathrm{M_{\odot}}$ & kpc & \\
         \hline
         m1e10 & $1\times10^{10}$ & $6.6\times10^{6}$ & $0.41$ &  $0.90$ & $0.0$\\
         m3e10 & $3\times10^{10}$ & $8.9\times10^{7}$ & $0.82$ & $0.77$ & $0.007$\\
         m1e11 & $1\times10^{11}$ & $1.4\times10^{9}$ & $1.68$ & $0.49$ & $0.20$\\
         m3e11 & $3\times10^{11}$ & $1.1\times10^{10}$ & $2.66$ & $0.30$ & $0.43$\\
         m3e11\_lowGas & $3\times10^{11}$ & $1.1\times10^{10}$ & $2.66$ & $0.10$ & $0.43$\\
         m3e11\_hiGas & $3\times10^{11}$ & $1.1\times10^{10}$ & $2.66$ & $0.50$ & $0.43$\\
         m1e12 & $1\times10^{12}$ & $3.1\times10^{10}$ & $3.10$ & $0.19$ & $0.66$\\
         \hline
    \end{tabular}
    \caption{Relevant initial conditions of each of the simulated galaxies. The table contains total galaxy halo mass, total stellar mass, exponential radius of stellar/gas disk, gas fraction of the galaxy and the bulge to total ratio (B/T).}
    \label{galaxy props}
\end{table*}
\subsection{Non-equilibrium chemistry of ions and molecules}
\label{chimes}
The chemistry of the gas within the simulated galaxies is modelled using \textsc{chimes}, a non-equilibrium chemistry and cooling module which can track 157 ions and molecules relevant for galaxy evolution \citep{richings2014a, richings2014b}. \citet{richings2014a} contains a full list of the \textsc{chimes} chemical reactions which includes all prominent processes in galactic chemistry. The reactions covering photochemistry require UV intensity which is calculated using $z=0$ UV background found in \citet{fgca2020} and the local radiation from star particles based on the \textsc{LEBRON} approximation \citep{hopkins2018a} and a treatment for self shielding (see \citet{richings2022} for detailed explanation of the UV treatment). HII regions are modelled by identifying neighbouring gas particles within a Stromgen radius of a given star particle, and disabling self-shielding for these gas particles so the ionisation is evolved in non-equilibrium with the full stellar flux provided from the star. These HII regions are of particular importance for the emission lines of ionised gas, such as H$\alpha$ and $[\mathrm{O}\textsc{iii}]_{5007\text{\AA}}$. The HII regions in our simulations have sizes varying between $\approx$5 - 100 pc, which is comparable to the spatial resolution of the gas at densities typical of HII regions. This means we can resolve individual HII regions however we cannot resolve the substructure within them.

A constant $\mathrm{H_{I}}$ ionization rate due to cosmic rays is used for all galaxies in this work ($\zeta_{\mathrm{HI}}=1.8\times10^{-16}\text{s}^{-1}$) and this is equivalent to the observed value in the Milky Way \citep{indriolo2012}. Ionization and disassociation rates of other species due to cosmic rays are then scaled relative to this value \citep{richings2022, richings2014a}. The high mass galaxies within our simulations have star formation rates comparable to the Milky Way however this means we overestimate the cosmic ray rate in our smaller dwarf galaxies, which will influence our predictions for the molecular phase where cosmic rays play a role in the chemistry. Dust surface reactions and dust shielding use an empirical density- and temperature-dependent model for dust abundance \citep{richings2022}, based on metal depletion observations \citep{jenkins2009, decia2016}.

\subsection{\textsc{RADMC-3D} and synthetic observations of emission line tracers}
\label{radmc3d}
Emission lines for a given species are generated for each simulation snapshot by post-processing the output data using \textsc{RADMC-3D}, a publicly available radiative transfer code which can track emission, absorption and propagation of spectral lines alongside emission from stellar sources and scattering and thermal emission from dust grains \citep{dullemond2012}.
Similar to \cite{richings2022}, the simulation output is projected face-on onto an adaptive mesh refinement (AMR) grid. Following that approach, each grid cell is refined until it contains, at most, 8 gas or star particles. Ion and molecular abundances, as well as species-weighted gas velocities and temperatures, are projected onto the AMR grid. Smoothing between cells uses the same cubic spline kernel as the MFM hydro solver. 

As in \cite{richings2022}, silicate and graphite grains are present in calculations run by \textsc{RADMC-3D}. The abundance of these grains at solar metallicity is taken from version 13.01 of \textsc{CLOUDY} \citep{mathis1977, ferland2013}, which is analogous to the ISM of the Milky Way. These abundances are then scaled by $Z/Z_{\odot}$ and then scaled again by the dust-to-metal (DTM) mass ratio \citep{richings2022}.  We run \textsc{RADMC-3D} for both the total emission of the galaxy and the continuum emission, before subtracting the continuum emission from the total emission. This leaves the emission from the galaxy disk for each emission line tracer.

The level populations for each observed species across each cell is calculated by \textsc{RADMC-3D}, which uses the non-equilibrium abundance of a given species and the gas properties within the cell. For the forbidden lines [C\textsc{ii}]$_{158\rm{\mu m}}$ and $[\mathrm{O}\textsc{iii}]_{5007\text{\AA}}$ we calculate the level populations using an optically thin non-LTE method, as the self-absorption of these lines is expected to be negligible. For CO$_{J(1-0)}$ we use the local velocity gradient (LVG) method, which accounts for local self-absorption of the CO line. For H$\alpha$ we cannot use \textsc{RADMC-3D} to calculate the level populations as it does not account for recombination radiation. Therefore, we calculate the contribution of HII recombinations to the Ha emissivity using the model from Liu et al. (in prep), which considers direct recombination radiation and the collisional excitation contribution from the ground state.

Atomic data and collisional excitation rates are taken from LAMBDA and CHIANTI respectively \citep{schoier2005, landi2013}. Line emissivity, in each cell, for each species is then calculated using the level populations. For each species' emission, a 3D position-position-frequency datacube is produced as a $\textsc{RADMC-3D}$ output which is then analysed using python. Spatially the cube is $12R_{disc}$kpc across where each pixel in the cube covers $20$pc spanning a range of wavelengths corresponding to velocities in the range $-200$\,kms/s - +$200$\,km/s centred on the emission line of interest, with a spectral resolution of 2\,km/s. For this work we do not attempt to model instrumental effects, such as the Point Spread Function or instrumental noise, for a particular telescope. Instead we focus on using the underlying spectra predicted by our simulations to test assumptions that are commonly used in observations to estimate physical properties of outflows.  

\subsection{Measuring outflow properties from synthetic spectra}
\label{synth spec}
Spectra of the emission lines ([C\textsc{ii}]$_{158\rm{\mu m}}$, $\mathrm{CO}_{J(1-0)}$, H$\alpha$, $[\mathrm{O}\textsc{iii}]_{5007\text{\AA}}$, [SII]$_{6731\text{\AA}}$, [SII]$_{6716\text{\AA}}$) are created by taking the sum of the intensity across the entire position-position grid in each of the 201 frequency slices. This results in a spectrum of 201 values against the Doppler shifted velocity dispersion of the emitting gas. A gaussian model is fit to the emission spectra and the model has 2 components denoting broad and narrow, similar to observational works, and these are fit with various initial parameters until a maximum $R^{2}$ score is found and the fit is confirmed by inspection. In general, the narrow components trace the kinematics of the ISM within a galaxy whilst the broad components trace the behaviour of the outflowing gas. Using the broad and narrow components of the decomposed spectra, it is possible to derive a velocity for the outflows using the equation from \citet{romano2023} based on \citet{rupke2005b}:

\begin{equation}
    \frac{v_{out}} {\mathrm{km\ s^{-1}}} = \frac{\mathrm{FWHM}_{broad}}{2} + |v_{broad} - v_{narrow}| 
\label{v out}
\end{equation}

where $v_{broad}$ and $v_{narrow}$ are the velocity peak of each component and $\mathrm{FWHM}_{broad}$ is the full width half maximum of the broad Gaussian component.

The outflow extent, $\mathrm{R_{out}}$, is defined as the region encompassing $95\%$ of the total emission luminosity for a given tracer \citep{romano2023}, with the implicit assumption that the outflow has the same spatial extent as the total emission. Following this, the kinematics of an outflow can be derived by using:
\begin{equation}
    \frac{\dot{M}_{out}}{\mathrm{M_{\odot}\ yr^{-1}}} = \frac{v_{out} \times M_{out}}{\mathrm{R_{out}}}
\label{m out rate}
\end{equation}

also from \citet{romano2023} based on \citet{rupke2005a}, where a thin shell time-averaged outflow is assumed where ${M}_{out}$ is the mass of the outflowing gas. Following on from this the momentum and energy outflow rates can be calculated using:

\begin{equation}
    \frac{\dot{p}_{out}}{\mathrm{g\ cm\ s^{-2}}} = v_{out} \times \dot{M}_{out} = M_{out} \frac{v_{out}^2}{\mathrm{R_{out}}}
    \label{p out rate}
\end{equation}

\begin{equation}
    \frac{\dot{E}_{out}}{\mathrm{erg\ s^{-1}}} = \frac{1}{2} \dot{M}_{out} \times v_{out}^{2} = \frac{1}{2} M_{out} \times \frac{v_{out}^{3}}{\mathrm{R_{out}}} 
    \label{E out rate}
\end{equation}

where $\dot M_{out}$ is the mass outflow rate (how much gas, in $M_{\odot}$ is moving out from the disk in a given time, in this case per year), $\dot E_{out}$ and $\dot p_{out}$ are the rate of energy outflow and momentum outflow respectively. Each of these equations requires a value for $M_{out}$, which is calculated for each emission line. Observational estimates of outflow properties are sensitive to an assumed inclination of the galaxy disk as the gas predominantly flows along the minor axis. Here we only consider galaxies that are viewed face on, however in the future these simulations could also be used to test the effects of inclination on the derived outflow properties.

\subsection{Emission line tracers}
\label{tracers}
The gas within galactic outflows can be traced by several emission lines within each individual phase. This subsection will cover how an outflow mass is calculated for the emission tracers chosen for this work, and these have been chosen to provide good coverage of the phases within an outflow which were mentioned in section \ref{intro}.

\subsubsection{[C\textsc{ii}]$_{158\rm{\mu m}}$}
\label{CII}
[C\textsc{ii}]$_{158\rm{\mu m}}$, the fine structure transition of $C^{+}$ at $158\mathrm{\mu m}$, is a well-known indicator of star formation within galaxies \citep{delooze2014, carniani2018, berthemin2020}. C\textsc{ii} is the strongest emission line within star forming galaxies in the rest-frame far infrared (FIR) and as a major coolant of the neutral atomic gas in the ISM \citep{stacey1991} it can also act as a tracer for this phase in outflows. The majority of [C\textsc{ii}]$_{158\rm{\mu m}}$ emission is found in this neutral atomic gas near the photo-disassociation regions of young O and B stars \citep{hollenbach1999, romano2023}. We calculate the outflow mass from the [C\textsc{ii}]$_{158\rm{\mu m}}$ luminosity ($L_{C\textsc{ii}}$) using the same equation that was used in \citet{romano2023} as follows:
\begin{multline}
    \frac{M^{atomic}_{out}}{\mathrm{M_{\odot}}} = 0.77 \left(\frac{0.7L_{C\textsc{ii}}}{L_{\odot}}\right) \left(\frac{1.4\times10^{-4}}{X_{C^{+}}}\right) \\ \times \left(\frac{1+2e^{-91K/T}+n_{\text{crit}}/n_{\text{e,gas}}}{2e^{-91K/T}}\right)
\label{CII luminosity}
\end{multline}

where $L_{C\textsc{ii}}$ is the luminosity in the broad C\textsc{ii} emission, $n_{\text{e,gas}}$ is the electron number density in the C\textsc{ii} line emitting gas, $n_{\text{crit}}$ is the critical density ($3\times10^{3}\mathrm{cm}^{-3}$) and $T$ is the [C\textsc{ii}]$_{158\rm{\mu m}}$ gas temperature, here assumed to be $130$\,K. This is the median value for gas temperatures found in \citet{romano2023} and is used in section \ref{synth spec vs observations}. In section \ref{testing assumptions} we test the impact of this assumption on outflow properties when compared to using the physical gas temperatures taken directly from the simulations. $X_{C+}$ is the ionised carbon abundance per hydrogen atom, initially set to $1\times10^{-4}$ in \citet{romano2023} following \citet{SandS1996}. This value assumes solar metallicity which does not hold for the galaxies in this work, which have different initial metallicities \citep{richings2022}. To remedy this we take the mass weighted averages of the metallicities in each snapshot, and assume that the carbon within the gas is all in C+, therefore we can use the carbon abundance for $X_{C+}$ which is a more accurate value than the initial assumption. $n_{\text{e, gas}}$ is calculated using the [SII]$_{6716\text{\AA}}$ and [SII]$_{6731\text{\AA}}$ lines. Utilising the luminosity derived from the broad emission component, across the outflow radius, of each of these lines we use:
\begin{equation}
    \frac{n_{\text{e}}}{\mathrm{cm^{-3}}} = \frac{cR_{broad} - ab}{a - R_{broad}}
\end{equation}
\label{n_e}
where $n_{\text{e}}$ is the electron density of the galaxy, $a=0.4315, b=2107, c=627.1, R_{broad}=\frac{L_{broad}^{SII6716}}{L_{broad}^{SII6731}}$, and the equation and constants are taken from \citet{sanders2016}. This equation, and the ratio between the two luminosity values is sensitive in the region of $50\text{cm}^{-3}<n_{\text{e}}<5000\text{cm}^{-3}$ \citep{fluetsch2021} and outside of this range the electron density does not vary significantly. \citet{sanders2016} derives the electron density from the SII line ratios with an assumed electron temperature of $10^{4}$\,K as this is a representative temperature for galaxies which are neither metal rich or metal poor and therefore it may overestimate and underestimate the respective electron densities, however the values are still within their calculated uncertainties. We use this method for calculating $n_{\text{e}}$ as it is the same method used by \citet{romano2023}, as opposed to alternative measures such as the [OII] doublet \citep{oii_ne}.

\subsubsection{$\mathrm{CO}_{J(1-0)}$}
\label{COJ10}
The majority of the mass in the molecular phase of the ISM and outflowing gas is found in molecular hydrogen. Ubiquitous in the Universe, molecular gas, traced by $H_{2}$, is the fuel behind star formation. Unfortunately, cold $H_{2}$ is all but invisible in emission, and instead a proxy must be used \citep{bolatto2013}. Commonly used is CO, with a lower critical density and excitation energy $(\frac{h \nu}{\kappa} \approx 5.53\,\mathrm{K})$ which makes it easy to excite in cold molecular clouds. The $CO(J=1-0)$ transition falls into a rather clear observational window ($\lambda = 2.6mm$) thus it can trace the molecular gas within galaxies far more easily than $\mathrm{H_{2}}$. Using a $\mathrm{CO}-\mathrm{H}_{2}$ conversion factor taken observations, $4.3\,\mathrm{\,\mathrm{M_{\odot} (K km s^{-1} pc^{2})^{-1}}}$, and the molecular gas mass equation from \citet{bolatto2013} the outflowing molecular gas mass can be calculated as follows:
\begin{equation}
    \frac{M_{out}^{mol}}{\mathrm{M_{\odot}}} = \alpha_{\text{CO}} L_{\text{CO}, broad}
\end{equation}
\label{mol m out}
where $\alpha_{\text{CO}}$ is the $\mathrm{CO}-\mathrm{H}_{2}$ conversion factor in $\,\mathrm{M_{\odot} (\ K\ km\ s^{-1}\ pc^{2})^{-1}}$. This equation leads to a total molecular gas mass within an outflow, including contributions from $\text{H}_2$ and He in line with observations. We correct for the presence of He here by multiplying our outflow masses by 1.36. In section \ref{testing assumptions} we test the impact of $\alpha_{\text{CO}}$ by comparing $4.3\,\mathrm{M_{\odot} (\ K\ km\ s^{-1}\ pc^{2})^{-1}}$ to values derived in \citet{accurso2017} which derives a multivariate $\alpha_{\text{CO}}$ factor that has been applied, for example, in the xCOLD GASS survey \citep{saintonge2017}, and which has been applied to our simulated galaxies in \citet{oliver2024} alongside values measured directly from the simulations in from the same paper and $0.8\,\mathrm{M_{\odot} (\ K\ km\ s^{-1}\ pc^{2})^{-1}}$, the $\alpha_{\text{CO}}$ value commonly used for ULIRGs and some of our observed galaxies here.
\subsubsection{$[\mathrm{O}\textsc{iii}]_{5007\text{\AA}}$}
\label{OIII5007A}
The $[\mathrm{O}\textsc{iii}]_{5007\text{\AA}}$ emission line is used as a tracer for ionised gas (see \citet{cicone2016} and references within). Young, hot stars ionise oxygen within the ISM producing the $[\mathrm{O}\textsc{iii}]_{5007\text{\AA}}$ line which traces warm ionised outflows. Utilising the electron density calculated previously from the SII doublet, assuming case B recombination \citep{osterbrock1989} and an assumed electron temperature of $T_{e}=10^4$K, the outflowing mass of $[\mathrm{O}\textsc{iii}]_{5007\text{\AA}}$ can be derived using the following equation from \citet{singha2022}.
\begin{equation}
    \frac{M_{out}^{ionised, OIII}}{\mathrm{M_{\odot}}} = (4\times10^{7}) \left(\frac{L_{OIII,broad}}{10^{43}}\right) \left(\frac{n_{\text{e}}}{100}\right)^{-1}
\end{equation}

\subsubsection{H$\alpha$}
\label{Halpha}
The H$\alpha$ is a strong optical emission line \citep{ceverino2016} and is another tracer of the ionised phase of gas outflows and is found around newly formed massive stars. Using an equation from \citet{wood2015} the broad component luminosity can be converted into an outflow mass by:
\begin{equation}
    \frac{M_{out}^{ionised, H\alpha}}{\mathrm{M_{\odot}}} = 6.4\times10^{6}\left(\frac{L_{H\alpha, broad}}{10^{42}\mathrm{ergs^{-1}}}\right) \left(\frac{500\mathrm{cm^{-3}}}{n_{\text{e}}}\right)
\end{equation}
where a gas temperature of $T_{gas} = 10^{4}$\,K is assumed. Case B recombination is also assumed.

\section{Outflow properties derived from synthetic emission line spectra}
\label{synthetic sepectra results}
\subsection{Comparison between simulation predictions and observational data}
\label{synth spec vs observations}
In this section, we present outflow properties ($\dot{M}_{out}$, $\dot{E}_{out}$, $\dot{p}_{out}$) derived from synthetic emission spectra of [C\textsc{ii}]$_{158\rm{\mu m}}$, H$\alpha$, CO$_{J(1-0)}$ and $[\mathrm{O}\textsc{iii}]_{5007\text{\AA}}$ using the techniques described in section \ref{synth spec} and compare them to observations. For each of the galaxy models we use five snapshots at intervals of 100\,Myr. Within the simulations, SFR is derived by taking the stellar mass formed averaged over the previous 10 Myr. \citet{richings2022} looked at the luminosities of a group of emission lines ([C\textsc{ii}]$_{158\rm{\mu m}}$, H$\alpha$, [O\textsc{i}]$_{63\rm{\mu m}}$, [O\textsc{iii}]$_{88\rm{\mu m}}$, [N\textsc{ii}]$_{122\rm{\mu m}}$) and find the simulations reproduce observed scaling relations with the SFR measured in the simulations.

\begin{figure*}
	\includegraphics[width=0.75\linewidth]{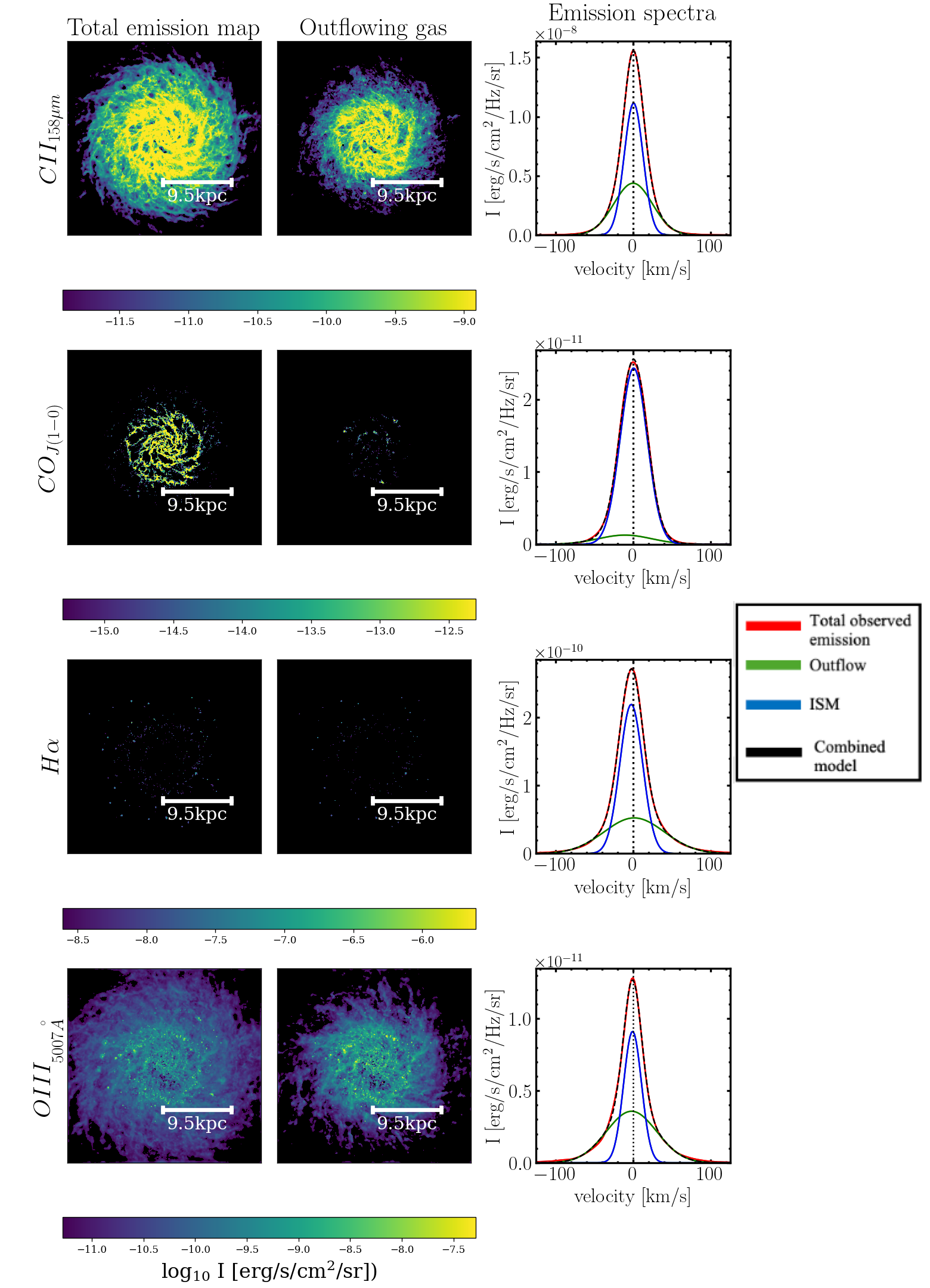}
    \caption{Emission maps and spectra for tracers of gas in m1e12, at 100\,Myrs. Left: total emission in each tracer for the whole galaxy. Centre: emission at velocities where the broad component is the dominant line. Cuts are based on the intercepts in the emission spectra components where the broad outflowing component dominates. Right: Emission spectra for each line and narrow (blue, ISM) and broad (green, outflow) components.The sum of the two components is given by the black dashed line, and the observed emission is the red line. CO is unique here as the broadest component denotes the ISM, as the gas appears to be in a stable disk without a visible outflowing component.}
    \label{m1e12 spectra}
\end{figure*}

Figure \ref{m1e12 spectra} shows emission maps and spectra for each of the emission lines from m1e12 during the simulation, after 100\,Myr. Each row represents a different emission line ([C\textsc{ii}]$_{158\rm{\mu m}}$, H$\alpha$, CO$_{J(1-0)}$ and $[\mathrm{O}\textsc{iii}]_{5007\text{\AA}}$ respectively). The left-hand column shows the total emission of the row's respective emission line, whilst the middle panel shows the emission from the broad, outflowing component. The right-hand panel shows the spectra with a best fit Gaussian made up of two components denoting the ISM and the outflow.

Outflows are defined using the spectra like those seen in figure \ref{m1e12 spectra}. For measuring the properties of an outflow, the entire broad component is considered. The luminosity of the emission is then calculated by integrating the area under the broad component curve. CO is converted into radio units before a luminosity is calculated in line with equation 6 from \citet{oliver2024}. The outflow properties gathered from the spectra in figure \ref{m1e12 spectra} are found in table \ref{m1e12 outflow props}.
 
\begin{table*}
    \centering
    \begin{tabular}{cccccc}
    \hline
         Line label & $L_{out}$ & $\mathrm{R_{out}}$ & $v_{out}$ & $M_{out}$ & $log_{10}(\dot{M}_{out})$\\
         & erg $\mathrm{s}^{-1}$ & kpc & km $\mathrm{s}^{-1}$ & $\text{M}_{\odot}$ & $\text{M}_{\odot}$ $\mathrm{yr}^{-1}$  \\
         \hline
         [C\textsc{ii}]$_{158\rm{\mu m}}$& $1.76\times10^{41}$ & $8.6$ & $29.8$ & $1.43\times10^{8}$ & $-0.796$\\
         $\mathrm{CO}_{J(1-0)}$ & $2.15\times10^{37}$ & $26.6$ & $52.0$ & $2.32\times10^{7}$ & $-1.83$\\
         $\mathrm{H\alpha}$ & $8.75\times10^{41}$ & $9.74$ & $49.7$ & $6.41\times10^{6}$ & $-1.97$\\         $\mathrm{[OIII]}_{5007\text{\AA}}$&$5.01\times10^{40}$&$11.8$&$39.5$&$4.59\times10^{4}$&$-4.30$\\
         \hline
    \end{tabular}
    \caption{Table collating observationally inferred outflow properties for m1e12, at 100\,Myrs. This contains data taken from the spectra in figure \ref{m1e12 spectra}, using the equations in section \ref{synth spec}.}
    \label{m1e12 outflow props}
\end{table*}

\begin{figure*}
	\includegraphics[width=0.75\linewidth]{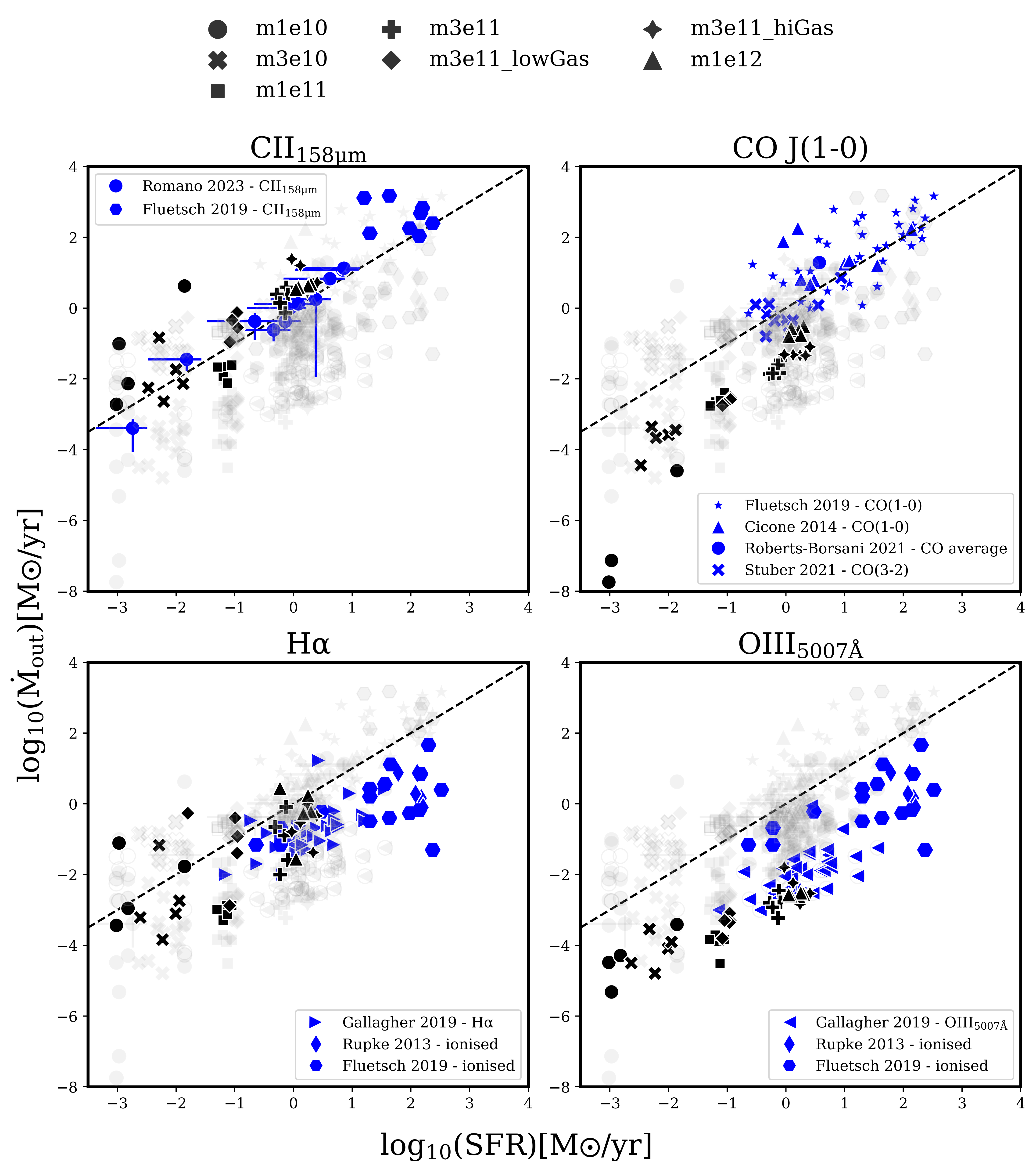}
    \caption{Observationally inferred mass outflow rate against SFR for each line for all simulation snapshots (black points), compared to observational data from the literature for the same emission line tracer (blue points), with grey points representing observations for the other phases \citep{romano2023, rupke2013, fluetsch2019, spilker2020, cicone2014, gallagher2019}. The simulated sample used in this work is in good agreement with trends seen in observations, especially in the neutral atomic phase traced by [C\textsc{ii}]$_{158\rm{\mu m}}$. The molecular phase tracer, $\text{CO}_{J(1-0)}$, is in a less good quantitative agreement ($\approx 1$dex difference at the same SFR) although the broad trends are still well followed. Observed outflow rates correspond to their original $\alpha_{CO}$ values, and the effects these have on our outflow rates are discussed in section \ref{CO-H2 test}. The dashed lines shows a mass loading factor of unity, where the mass outflow rate is equal to the SFR.}
    \label{mass outflow plots}
\end{figure*}
\begin{figure*}
	\includegraphics[width=0.75\linewidth]{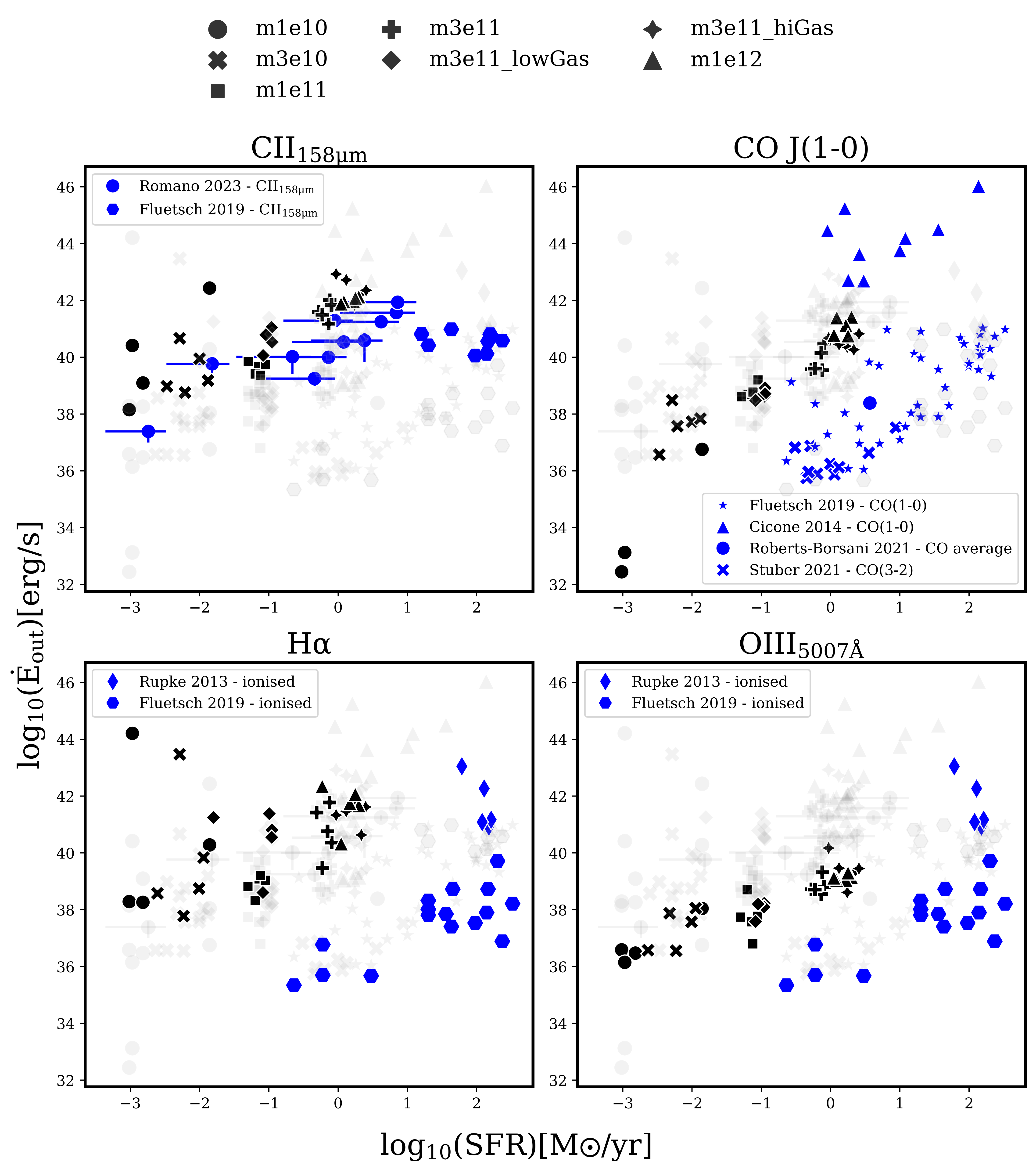}
    \caption{Observationally inferred energy outflow rate vs SFR for this work's simulated sample and a number of observational works. Similar to figure \ref{mass outflow plots}, the [C\textsc{ii}]$_{158\rm{\mu m}}$ emission is in better agreement, with the other lines following similar qualitative trends to the observations.}
    \label{E outflow plots}
\end{figure*}
\begin{figure*}
	\includegraphics[width=0.75\linewidth]{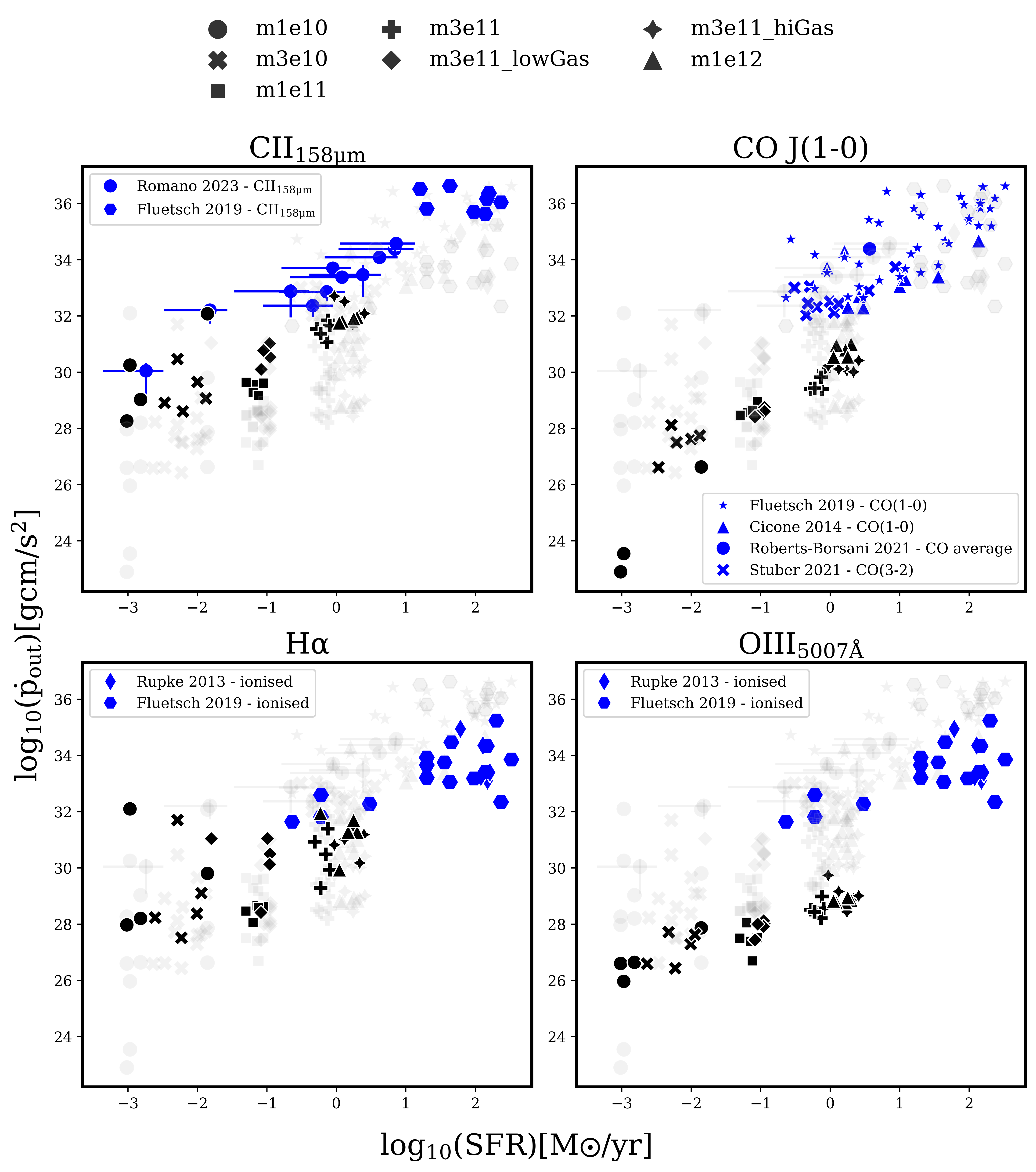}
    \caption{Observationally inferred momentum outflow rate vs SFR for the simulated sample and observations. As in the previous plots, the simulations are in qualitative agreement with observational data, following the broad trends well. However, the momentum flux is systematically under predicted in these tracers when compared to observations.}
    \label{p outflow plots}
\end{figure*}

Applying equations from section \ref{synth spec} for outflow properties across the entire sample of galaxies and snapshots, we can compare the properties derived in this work to observational work using the same emission line tracers. The first panel in figure \ref{mass outflow plots} is [C\textsc{ii}]$_{158\rm{\mu m}}$, in the top left. Here we compare to observations from \citet{romano2023} and star formation dominated galaxies from \citet{fluetsch2019}. The former of these investigates dwarf galaxies in a mass range of $10^{6}\,\text{M}_{\odot} - 10^{10}\,\text{M}_{\odot}$ and with metallicities of $<0.5 Z_{\odot}$, making them ideal for comparing to our lower mass haloes in this work. \citet{fluetsch2019} primarily focusses on molecular outflows, although analyses of the atomic phase are also included. They focus on ALMA data and their data is more biased towards starburst galaxies, which will be more active than the galaxies shown here although they remain a useful high-mass comparison. Both works observe local galaxies and we find that our outflow properties are in good agreement, especially with \citet{romano2023}, although we caveat this with the fact we applied similar assumptions to their work and we will test these assumptions in sections \ref{CII temp test} and \ref{n_e assumption test}.

Next, in the top right we have $\mathrm{CO}_{J(1-0)}$ and we compare this to observations of the $\mathrm{CO}_{J(1-0)}$ line from \citet{fluetsch2019}, the star formation dominated galaxies from \citet{cicone2014}, observations of the $\mathrm{CO}_{J(2-1)}$ line from \citet{stuber2021} and average $\mathrm{CO}_{J(1-0)}$ properties across a combination of large scale surveys from \citet{rb2020}. \citet{fluetsch2019} and \citet{cicone2014} are each biased towards starbursts and as such use $\alpha_{\text{CO}}$ values of $0.8\,\mathrm{M_{\odot} (\ K\ km\ s^{-1}\ pc^{2})^{-1}}$ whilst \citet{stuber2021} uses the Milky Way-like value of $4.3\,\mathrm{M_{\odot} (\ K\ km\ s^{-1}\ pc^{2})^{-1}}$, which we apply to our galaxies here, and we will test the differences in these values in section \ref{CO-H2 test}. Here we find a larger difference between our galaxies and the observations, alongside a weaker relation between SFR and outflow rate, with the latter increasing by around 1 dex over a SFR range of 7 dex. We find a roughly 1 dex underprediciton of mass outflow rates for our highest mass galaxies when compared to the observations. 

The $\mathrm{H\alpha}$ (bottom left) and $[\mathrm{O}\textsc{iii}]_{5007\text{\AA}}$ (bottom right) lines are compared to \citet{fluetsch2019} and the low-redshift observations from \citet{gallagher2019} and low-redshift starburst mergers from \citet{rupke2013}. We find the two ionised lines are in good agreement with observations at similar SFRs, especially for our Milky Way mass snapshots. The H$\alpha$ line has a larger scatter in outflow rates, with $\dot{M}_{out}$ changing by 4 dex across our range of galaxy masses, and a difference in about 2 dex between our low mass galaxies (m1e10, m3e10) and the observational samples. $[\mathrm{O}\textsc{iii}]_{5007\text{\AA}}$ shows similar behaviour with less scatter in $\dot{M}_{out}$. 

Overall, the strongest outflows are seen in the Milky Way mass galaxy in the [C\textsc{ii}]$_{158\rm{\mu m}}$ line, with mass outflow rates of $~0.1\mathrm{M_{\odot}yr^{-1}}$, whilst the weakest outflows are found in the dwarf mass galaxy (m1e10) in $\mathrm{CO}_{J(1-0)}$ emission with mass outflow rates ranging between $0.01 - 1\times10^{-6} \mathrm{M_{\odot}yr^{-1}}$ across the five simulation snapshots.

The energy outflow rates ($\dot{E}_{out}$, figure \ref{E outflow plots}) show a wider spread of outflow rates for both the simulations and the observations. This is especially apparent in $\mathrm{CO}_{J(1-0)}$, which differs by up to $\approx10$ dex in at a fixed SFR in observations whilst the simulations shows a tighter relationship. In terms of the outflow rates themselves, the ionised lines appear to over predict $\dot{E}_{out}$ by around 2-4 dex relative to observations at the same SFR. $\mathrm{CO}_{J(1-0)}$ lies in the middle of the previously mentioned observational scatter, whilst [C\textsc{ii}]$_{158\rm{\mu m}}$ still appears in relatively good agreement with observations at matching SFRs. $\mathrm{CO}_{J(1-0)}$ and $[\mathrm{O}\textsc{iii}]_{5007\text{\AA}}$ also both under predict the momentum flux ($\dot{p}_{out}$, figure \ref{p outflow plots}) by around 2 dex, whilst [C\textsc{ii}]$_{158\rm{\mu m}}$ and $\mathrm{H\alpha}$ show smaller differences of about 0.5 dex to observations at the same SFR. This highlights potential discrepancies in the outflow velocities, but we leave this work for the future. The bursty nature of star formation may be behind these differences as high star formation in a concentrated region will drive more powerful outflows relative to weaker star formation. This is especially pronounced in the dwarf mass galaxies, where bursty star formation can actually drive the entire reservoir of gas out of the galaxy. The weakest outflows are also in m1e10, this time in the $[\mathrm{O}\textsc{iii}]_{5007\text{\AA}}$ emission, which has $\dot{E}_{out}\sim10^{34}$ erg/s alongside a smaller SFR than observed galaxies with similar energy outflow rates. 

The momentum outflow rates ($\dot{p}_{out}$, figure \ref{p outflow plots}) again follow the trend of increasing in line with SFR, although now there is a larger discrepancy between simulations and observations across all emission lines. For example, in the [C\textsc{ii}]$_{158\rm{\mu m}}$ emission, at a fixed SFR the simulations predict momentum fluxes which are $~1-2$ dex lower than observations. The highest-momentum outflow is found in m3e11\_hiGas in the [C\textsc{ii}]$_{158\rm{\mu m}}$ line with a value of $\dot{p}_{out}\sim10^{33}\,\mathrm{gcms^{2}}$ whilst the lowest is in the CO$_{J(1-0)}$ line at $\dot{p}_{out}\sim10^{23} \,\mathrm{gcms^{2}}$. 

When comparing the mass outflow rates, energy outflow rates and momentum outflow rates with each other, the best fits are consistently found in the [C\textsc{ii}]$_{158\rm{\mu m}}$ line, followed by the two ionised lines, $\mathrm{H\alpha}$ and $[\mathrm{O}\textsc{iii}]_{5007\text{\AA}}$ and then CO$_{J(1-0)}$. The mass outflow rates show the best agreement between simulated results and observations and the clearest trends in the synthetic data, followed by the momentum outflow rates and then the energy outflow rates.

The plots show that mock observations of the simulated galaxies can recreate observed trends and outflow properties. Following this we can use the simulations to test some common assumptions which go into the derivation of outflow properties. Then we can compare outflows as defined by simulators, and see if simple comparisons between simulations and observations are applicable despite differing methods.

\subsection{Phase-space properties of the line emitting gas}
\label{phase-space}
\begin{figure*}
	\includegraphics[width=0.65\linewidth]{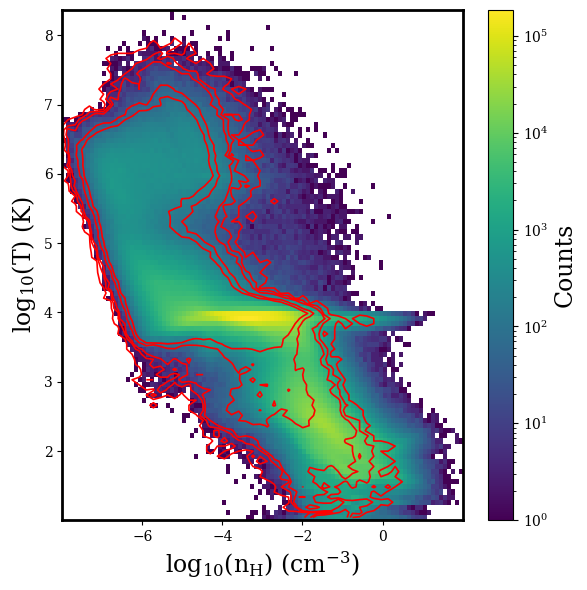}
    \caption{Temperature-density histogram for m1e12 at 100\,Myrs. This is made using the particles from the simulation snapshot, and counts represents the number of particles in a given temperature-density bin. The overplotted contours represent the temperature-density histogram with a velocity cut to isolate the outflowing gas. In this case the velocity cut is set to the fastest outflow velocity (52.0 km $\mathrm{s}^{-1}$) taken from table \ref{m1e12 outflow props}.}
    \label{TnH diagram}
\end{figure*}

\begin{figure*}
	\includegraphics[width=0.75\linewidth]{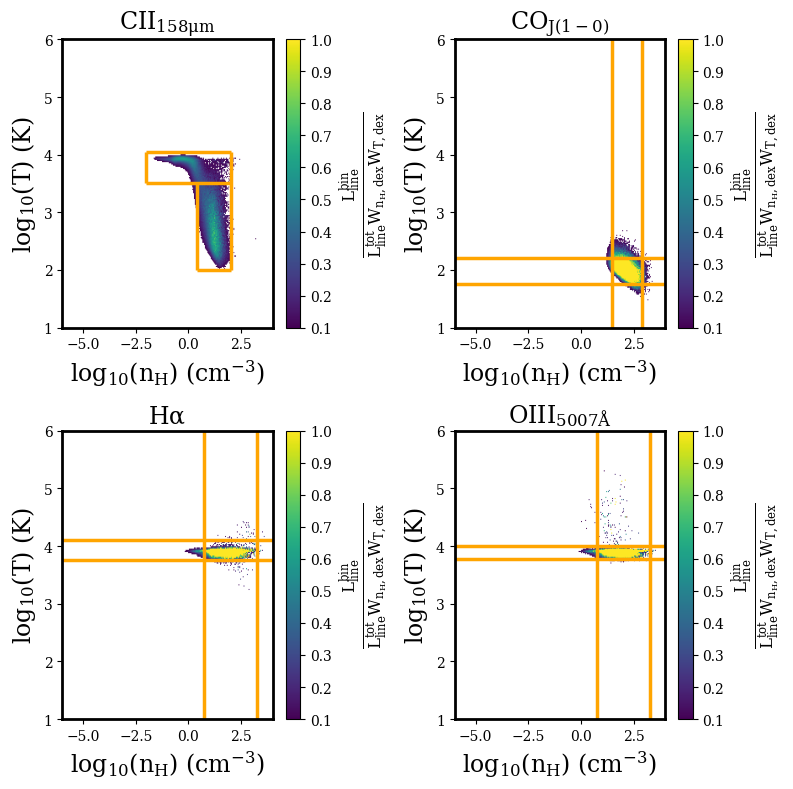}
    \caption{Temperature density histograms for m1e12 at 100\,Myrs. Top left: [C\textsc{ii}]$_{158\rm{\mu m}}$, top right: CO$_{J(1-0)}$, bottom left: H$\alpha$, $[\mathrm{O}\textsc{iii}]_{5007\text{\AA}}$. Points are coloured by $\frac{L^{bin}_{line}} {L^{tot}_{line} W_{nH, dex} W_{T, dex}}$ with a minimum value of 0.1 imposed, where $L_{bin}^{line}$ is the line luminosity for a given bin, $L_{bin}^{tot}$ is the total luminosity for all lines in a given bin (not just those investigated in this work), $W_{nH, dex}$ is the width of the density bins in dex and $W_{T, dex}$ is the width of the temperature bins in dex. Similar to figure \ref{TnH diagram}, a velocity cut is applied to ensure we are only looking at gas moving at a line of sight velocity greater than 52 $\mathrm{km\,s}$.}The brightest points within this space are highlighted by the orange boxes. These bright n-T regions within these orange boxes are then defined as the particles corresponding to a given phase. For example, the bright region in CO$_{J(1-0)}$ (top right) are used to define the cold molecular phase in section \ref{outflow props from particles}.
    \label{m1e12 T-n cuts}
\end{figure*}

Figure \ref{TnH diagram} compares the temperature-density distribution of ISM gas (shown in the coloured histogram) to the temperature-density distribution of the outflowing gas, which we define as gas with an absolute line of sight velocity greater than 52 $\mathrm{km\,s}$, which is the fastest outflow velocity within our observed lines at in m1e12 100 $\mathrm{Myr}$ in table \ref{m1e12 outflow props} (outflowing gas is shown by the contours). Overall the temperature-density distribution within the outflowing gas is very similar to that of the ISM, and this is influenced by the metal cooling. We would expect differences in the temperature-density distributions if the outflowing gas and the ISM had different metallicities, so the similarities between them implies that by extension they have similar metallicities. Generally, we would expect outflows to be enriched following the impact of stellar feedback, but m1e12 already starts off with a high metallicity and the overall metallicity of the galaxy does not evolve much over the course of the simulation, which suggests that the enrichment from new star formation is small compared to the metallicity we started with in the initial conditions, consistent with what we see in the plot. Unlike the ISM, the outflowing gas does not extend to higher densities which suggests that the outflows destroy pre-existing molecular clouds as they sweep up the gas, although a deeper investigation of this is beyond the scope of this work.

Figure \ref{m1e12 T-n cuts} explores the temperatures and densities of the outflowing gas (with outflows defined in the same way as for Figure \ref{TnH diagram}) which produces each of the emission lines covered in the previous section, to get a better understanding of what gas each line is actually tracing, in this case for m1e12 at 100\,Myrs. Each bin on the figure is the emission intensity normalised by the total luminosity of the emission line (over the whole AMR grid) and by the width of each temperature and density bin in dex. The bulk of the emission for the gas particles in each phase is isolated and these can be seen by the orange boxes surrounding the brightest areas in Figure \ref{m1e12 T-n cuts}. [C\textsc{ii}]$_{158\rm{\mu m}}$ is an exception here as it is found over a greater range of temperatures and densities, meaning two cuts are required to capture the total emission. These temperature-density values within the boxes are then used to derive outflow properties for a specific phase of gas, corresponding to our emission line tracers, in section \ref{pixel motivated}.

The behaviour of the gas in each phase is as we expect. The [C\textsc{ii}]$_{158\rm{\mu m}}$ emitting gas is more diffuse than the other gas phases and is the majority of what we see throughout the outflow. The cold molecular gas is both the densest ($1.48<\mathrm{log}(n_{\mathrm{H}}[\text{cm}^{-3}])<2.88$) and the coldest ($1.76<\mathrm{log}(T[\text{K}])<2.20$). The tracers of the hot ionised phase, $\mathrm{H\alpha}$ and $[\mathrm{O}\textsc{iii}]_{5007\text{\AA}}$, both trace a similar region in the phase-space diagram, with peak temperatures $~10^{4}$\,K and larger distribution of densities for the bulk of the gas ($\mathrm{log}(n_{\mathrm{H}}[\text{cm}^{-3}])\simeq 1-3$) that generally favours lower values when compared to the molecular phase. These phase-space cuts will allow us to trace gas outflows corresponding to the previously observed phases in section \ref{outflow props from particles}.

As the simulations are Lagrangian, the mass resolutions are fixed, but the effective spatial resolutions of the gas varies with density. Figure \ref{m1e12 T-n cuts} shows that different tracers probe different densities thus they also probe the gas at different resolutions. [C\textsc{ii}]$_{158\rm{\mu m}}$ traces is the most abundant of the tracers used in this work, and covers a range from of phases from cool to warm atomic gas. Therefore it has the broadest resolution range, between 10 - 206 pc. $\mathrm{CO}_{J(1-0)}$ traces molecular gas at relatively high densities and so resolves gas at spatial scales of 5 - 14 pc. $\mathrm{H\alpha}$ and $\mathrm{[OIII]}_{5007\text{\AA}}$ both trace warm ionised gas at those same relatively high densities and therefore resolve gas at spatial scales of 4 - 25 pc.

\subsection{Testing observational assumptions for measuring outflow properties}
\label{testing assumptions}

\subsubsection{[C\textsc{ii}]$_{158\rm{\mu m}}$ temperature}
\label{CII temp test}
In section \ref{synth spec vs observations} the outflow properties traced by [C\textsc{ii}]$_{158\rm{\mu m}}$ are calculated using a gas temperature of 130\,K, which is used to match the assumptions made in observations (primarily in \citealt{romano2023}). However in section \ref{phase-space} we saw that [C\textsc{ii}]$_{158\rm{\mu m}}$ emission arises across a wide range of gas temperatures (see figure \ref{m1e12 T-n cuts}). In this section we will use the simulations to test how an assumed temperature of [C\textsc{ii}]$_{158\rm{\mu m}}$ affects our outflow properties. We do this by calculating a [C\textsc{ii}]$_{158\rm{\mu m}}$-weighted mean temperature in each pixel of the [C\textsc{ii}]$_{158\rm{\mu m}}$ emission map, averaging across all AMR cells along the line of sight which are contributing to a given pixel. The average temperature within the outflow radius R$_{out}$ is then taken, also averaged by [C\textsc{ii}]$_{158\rm{\mu m}}$. 
\begin{figure}
	\includegraphics[width=\columnwidth]{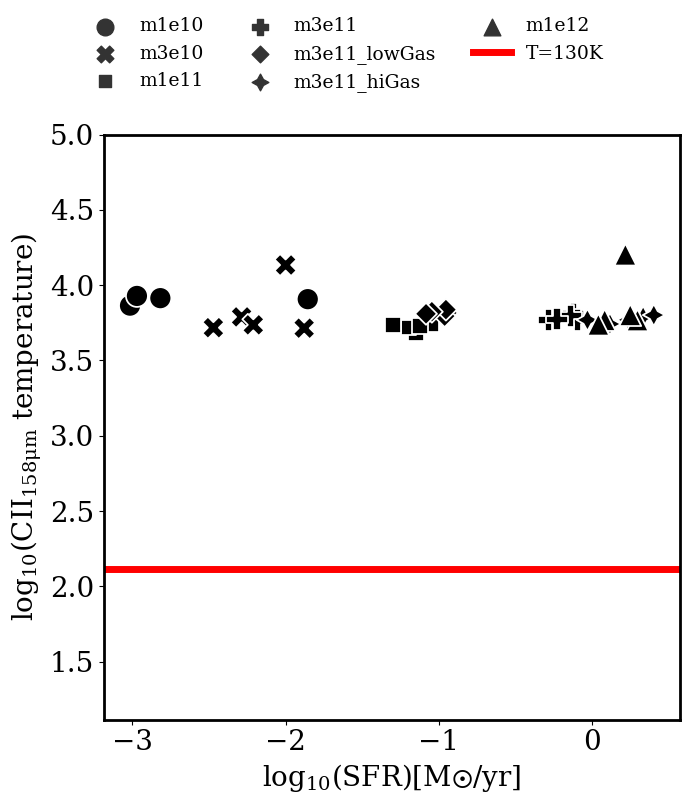}
    \caption{[C\textsc{ii}]$_{158\rm{\mu m}}$-weighted mean gas temperature against SFR for all simulation snapshots. Here we see a decrease in temperature across the SFR range, of around $\sim 0.2$ dex. We can also see that the actual measured temperatures of our outflowing gas is higher than the initially assumed value of 130\,K.}
    \label{CII temps vs SFR}
\end{figure}
Figure \ref{CII temps vs SFR} shows a weak relation between the average temperature within R$_{out}$ and the SFR of the host galaxy. We find that our measured temperatures are closer to those seen in the \citet{sanders2016} electron density equations rather than the 130\,K used as a median seen in \citet{romano2023}. However, quite a few of our galaxies are still up to 0.3 dex lower than the \citet{sanders2016} temperature, and so this might still contribute to the uncertainties in the electron density in the next section. We find that the average temperature decreases by roughly 0.2 dex across a 3 dex increase in SFR. It also shows that the actual particle temperatures are far higher than the assumed value of 130\,K, by more than an order of magnitude. 

\begin{figure}
	\includegraphics[width=\columnwidth]{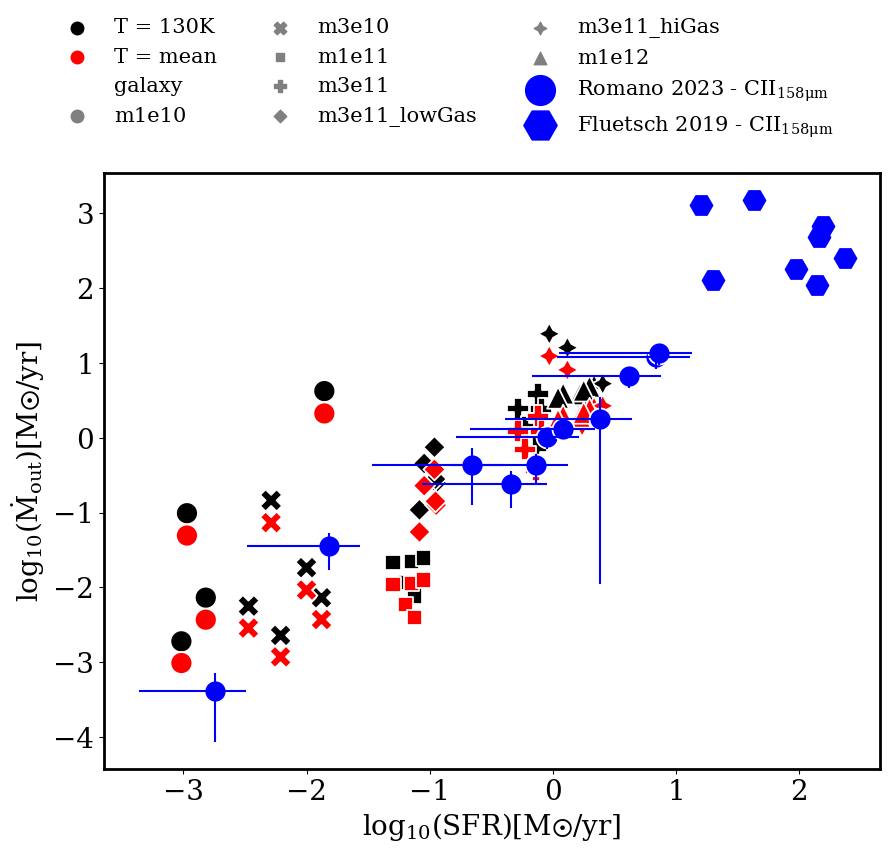}
    \caption{Mass outflow rate against SFR for [C\textsc{ii}]$_{158\rm{\mu m}}$ for all simulation snapshots (black points), compared to observational data from the literature for the same emission line tracers (blue points) as in the [C\textsc{ii}]$_{158\rm{\mu m}}$ panel of figure \ref{mass outflow plots}. Red: [C\textsc{ii}]$_{158\rm{\mu m}}$ mass outflow rates calculated using the [C\textsc{ii}]$_{158\rm{\mu m}}$-weighted mean gas temperature within the projected outflow radius in each simulation. We see very little variation in the mass outflow rates when using 130\,K or the measured values as our gas temperatures, because equation \ref{CII luminosity} is not sensitive when $\mathrm{T}\gg91\,\mathrm{K}$.}
    \label{CII temperatures outflows}
\end{figure}

In figure \ref{CII temperatures outflows} we recalculate mass outflow rates traced by [C\textsc{ii}]$_{158\rm{\mu m}}$ using the CII-averaged temperature (red points) and compare them to mass outflow rates using an assumed temperature of 130\,K (black points), alongside observational data (blue points). We find that using the CII-averaged temperatures has little effect on the observed mass outflow rate of our simulated galaxies. Where it does have a visible effect it, the mass outflow rate actually decreases, although by no more than ~0.2 dex.

Figure \ref{CII temperatures outflows} also shows that our measured [C\textsc{ii}]$_{158\rm{\mu m}}$ masses may not be very sensitive to the gas temperature, as it appears to have little impact on mass outflow rates. In equation \ref{CII luminosity} temperature appears in the term $\text{e}^{-91K/T}$ which is close to unity if $\mathrm{T}\gg91\mathrm{K}$, which it is for the directly measured temperatures in our galaxies.

\subsubsection{Electron density}
\label{n_e assumption test}
Electron density is used in the mass outflow rate equations for [C\textsc{ii}]$_{158\rm{\mu m}}$, H$\alpha$ and $[\mathrm{O}\textsc{iii}]_{5007\text{\AA}}$. In section \ref{synth spec vs observations} we calculated the electron density using the SII doublet ratio \citep{sanders2016}, to match observational methods. This method for measuring electron density is only sensitive in the region of $50<n_{e}\,[\mathrm{cm^{-3}}]<5000$ \citep{fluetsch2021}. Furthermore, figure \ref{m1e12 T-n cuts} shows that different emission lines trace gas at different densities, which may further bias outflow properties and as such may fit better to some of our emission line than others. We can test this by finding the mean electron density from the simulation data for each snapshot, and weighting it by each emission line. We can then see whether or not this has a strong effect on our inferred outflow properties and therefore decide if the SII doublet is applicable to our synthetic observations.

\begin{figure}
	\includegraphics[width=\columnwidth]{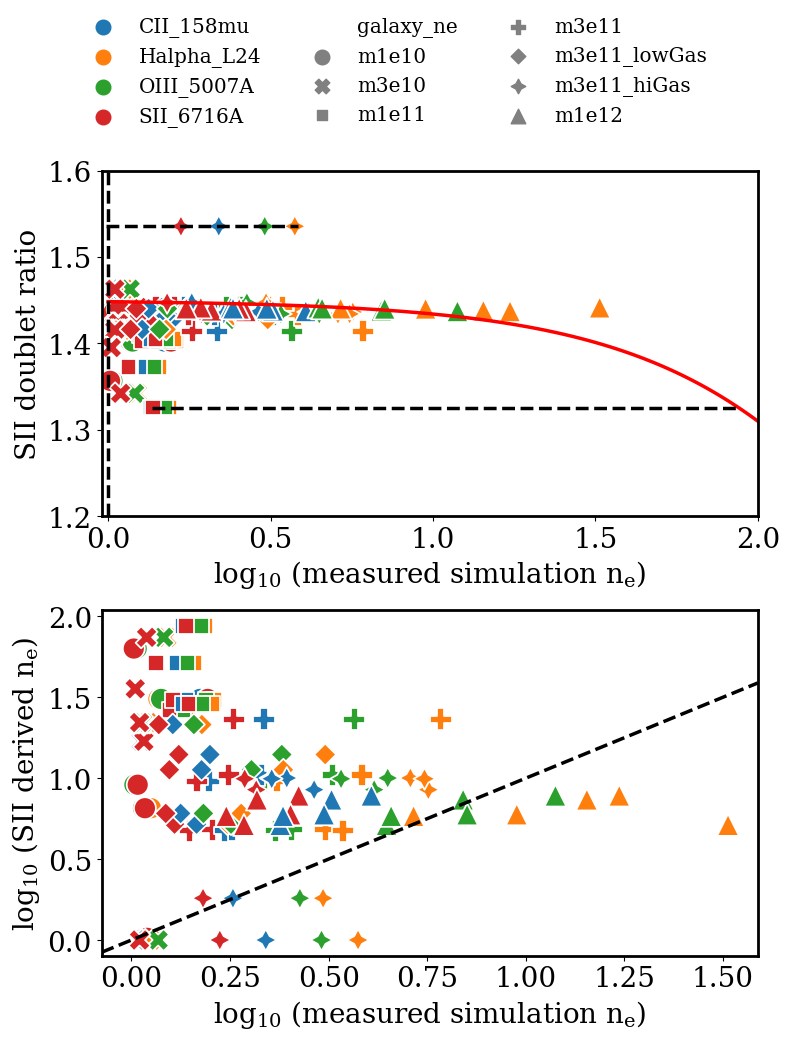}
    \caption{Top panel: SII doublet ratio against average luminosity weighted electron density for emission lines utilising $n_{\text{e}}$ for calculating outflow properties. Red line represents values from the relation in \citet{sanders2016}. The vertical dashed line shows the limit for electron density whilst the horizontal lines show where simulated points should theoretically sit on the \citet{sanders2016} relation. Bottom panel: electron density derived from the SII doublet vs average luminosity weighted electron density from the simulation. Here the black dashed line represents a 1:1 relation between the two methods, and the closer to this line a point lies the better the SII doublet is at approximating the actual electron density. Points lying along the bottom are set to 1 in line with \citet{sanders2016} and this is limit is described by the vertical dashed line. Horizontal dashed lines are used to show where the snapshots should lie following the \citet{sanders2016} relation. Our points lie along the upper limit described in \citet{sanders2016} with the SII doublet method returning higher electron densities than we measure directly.}
    \label{SII_ratios}
\end{figure}

The top panel of figure \ref{SII_ratios} shows the measured SII doublet ratios in our snapshots compared to the physical electron density derived from the particles within the simulation. The red line describes the relationship stated in \citet{sanders2016} and the figure shows that our measured SII ratios generally sit along the upper limit of the relationship found in \citet{sanders2016} and therefore fall out of the range of sensitivity for the relation ($50 < n_{e}\,[\mathrm{cm^{-3}}] < 5000$). Some of our points lie above the red line, whilst some points also lie below this line and in this case the SII doublet over predicts the electron density relative to the actual measured value, and this is represented by the dashed line towards the bottom of the plot. The bottom panel of figure \ref{SII_ratios} is then used to visualise how well the entire \citet{sanders2016} SII doublet relation predicts $n_{\text{e}}$ by comparing them to the actual values for $n_{\text{e}}$ as calculated from the simulations. From the plot, we see a large discrepancy between the SII doublet and direct measurements, especially for m1e12 which instead follows a horizontal line rather than the 1:1 relation denoted by the dashed line. An explanation for why the SII derived densities differ from the simulation-measured values could be that the SII lines are tracing different gas from the other lines, such as [C\textsc{ii}]$_{158\rm{\mu m}}$. However the SII lines themselves also deviate from the both the \citet{sanders2016} relation and simulation-measured electron densities. This could point towards a mixture of electron densities and temperatures along the line of sight.

\begin{figure}
	\includegraphics[width=\columnwidth]{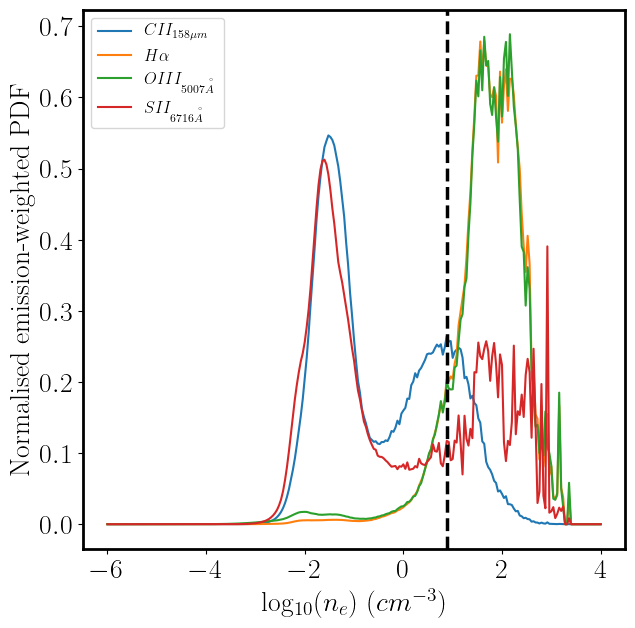}
    \caption{Histogram showing emission weighted electron densities for the relevant emission tracers used in this work within m1e12 at 100\,Myr. we find that the peaks correspond to where we would expect when compared to figure \ref{m1e12 T-n cuts}. We find that [C\textsc{ii}]$_{158\rm{\mu m}}$ peaks across a wider range of densities whilst $[\mathrm{O}\textsc{iii}]_{5007\text{\AA}}$ and H$\alpha$ share a similar peak at roughly $2\,\mathrm{cm^{-3}}$. $[\mathrm{S}\textsc{ii}]_{6716\text{\AA}}$ peaks at both $-2\,\mathrm{cm^{-3}}$ and $2\,\mathrm{cm^{-3}}$. The black dashed line is the electron density as measured using the SII doublet ratio and covers the weaker part of the cool atomic distribution but struggles with the other tracers.}
    \label{ne_pdf}
\end{figure}
Figure \ref{ne_pdf} provides a deeper insight into the electron densities of our simulated galaxies, and the observed emission tracers. The $[\mathrm{O}\textsc{iii}]_{5007\text{\AA}}$ line is prevalent in regions with electron densities around $100\, \mathrm{cm^{-3}}$ which corresponds to the typical densities of HII regions. The bimodal distribution of the CII and SII lines is likely due to the strength of these lines in both the cold and warm phases, with each peak likely corresponding to these phases.

\begin{figure}
	\includegraphics[width=\columnwidth]{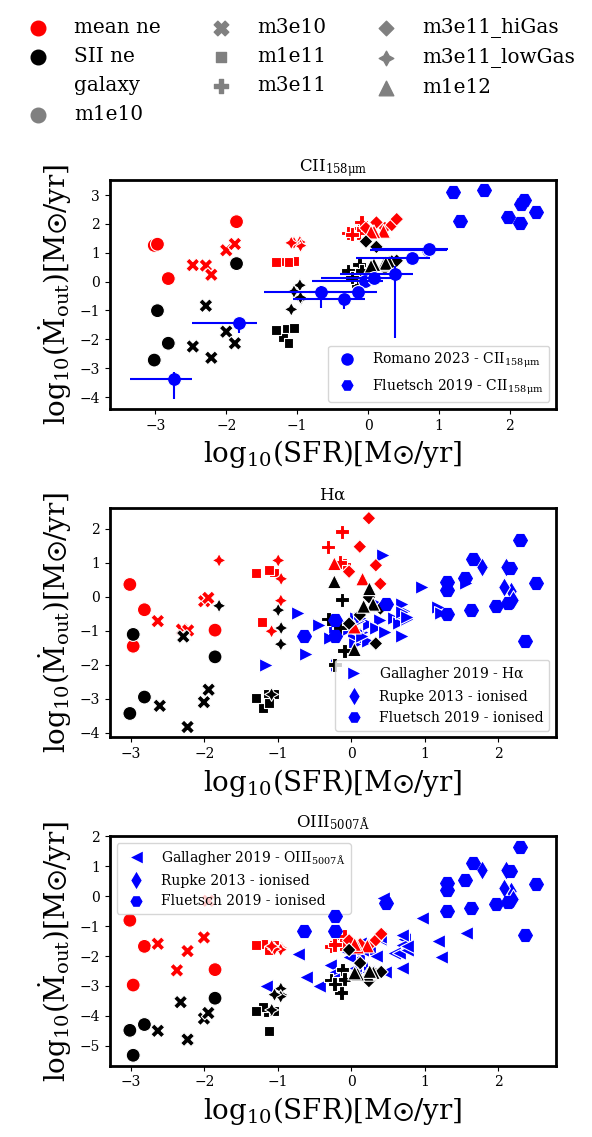}
    \caption{Mass outflow rates compared to SFR for a selection of simulated galaxies and observational results for the [C\textsc{ii}]$_{158\rm{\mu m}}$, H$\alpha$, and $[\mathrm{O}\textsc{iii}]_{5007\text{\AA}}$ emission lines. Blue points represent values from observational literature, black points are using the SII doublet ratio to estimate $\mathrm{n_{e}}$ (as used initially in figure \ref{mass outflow plots}) and red points are using the electron densities as measured directly from the simulations. We find that our outflow rates are sensitive to the electron density and that the choice of method impacts the outflow rates by up to ~4 dex.}
    \label{ne_outflow_tests}
\end{figure}

In figure \ref{ne_outflow_tests} we plot the mass outflow rates of the [C\textsc{ii}]$_{158\rm{\mu m}}$, H$\alpha$, and $[\mathrm{O}\textsc{iii}]_{5007\text{\AA}}$ lines using both the measured electron densities for each galaxies and the SII doublet electron densities to see how the observational methods affect our outflow kinematics.
Figure \ref{ne_outflow_tests} shows a difference of up to 4 dex when compared to using the SII ratios.  This large discrepancy shows that care must be employed when using the SII ratios as there is potentially large variation between inferred and actual electron densities within galaxies and outflows themselves. This difference, and the restrictive range at which the SII doublet is sensitive to electron density, imply that a different tracer may be more applicable for the study of outflows. \citet{kewey2019} provides a list of several alternative tracers across optical, UV and IR lines. Optically, there is [OII]$_{3727\text{\AA}}$ and [OII]$_{3729\text{\AA}}$ lines which are sensitive in the range $40<n_{e}\,[\mathrm{cm^{-3}}]<10000$ although within this work the SII doublet is preferred for consistency with the observations we are comparing to. Whist the simulations could potentially be used for debiasing observations with respect to electron density, figure \ref{SII_ratios} shows a large scatter between $\mathrm{n_{e}}$ measured observationally and $\mathrm{n_{e}}$ measured directly from the simulations. This means there is not a simple one-to-one relation between the two which would allow for debiasing. Using the SII doublet to derive an electron density in this way also results in an average value across the galaxy whilst, in practice, line emitting gas is likely to have a distribution of different electron densities.

\subsubsection{CO-to-H$_{2}$ conversion factor}
\label{CO-H2 test}
In section \ref{COJ10} we converted CO luminosity to $\mathrm{H}_{2}$ mass assuming a CO-to-$\mathrm{H}_{2}$ conversion factor, $\alpha_{\text{CO}}$, of $4.3\,\mathrm{M_{\odot} (\ K\ km\ s^{-1}\ pc^{2})^{-1}}$ which is used in observations and in the Milky Way \citep{bolatto2013}. This section will test the impact of this assumption on our outflow properties. We recalculate the mass outflow rates using an $\alpha_{\text{CO}}$ measured directly from the simulations \citep{oliver2024}, which in effect means we can directly utilise the actual $\mathrm{H}_{2}$ mass, and we also test an $\alpha_{\text{CO}}$ from \citet{accurso2017}.
\begin{figure}
	\includegraphics[width=\columnwidth]{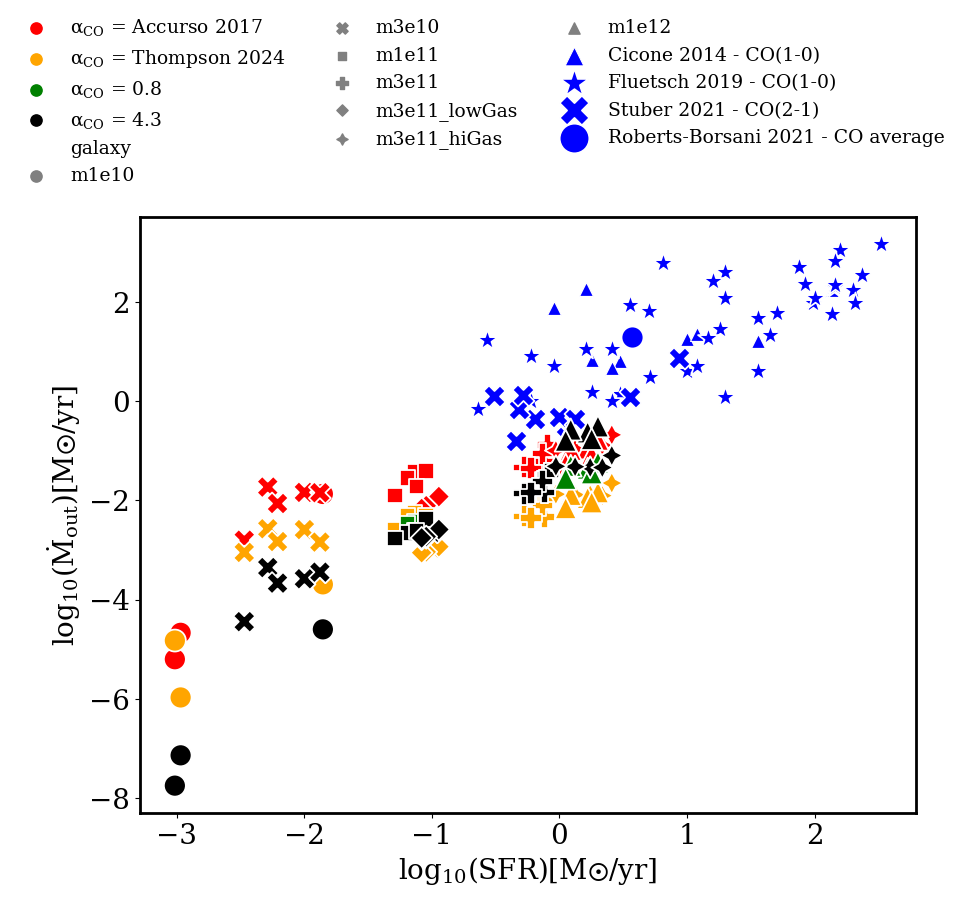}
    \caption{Mass outflow rate against SFR for CO, for all simulation snapshots (black points), compared to observational data from the literature for the same emission line tracers (blue points) as in the CO$_{J(1-0)}$ panel of figure \ref{mass outflow plots}. The red points represent the outflow rates calculated using the directly derived CO mass using $\alpha_{\text{CO}}$ from \citet{oliver2024}. We find that the choice of $\alpha_{\text{CO}}$ affects the outflow rates by 0.5 dex, with both observer based values increasing the mass outflow rates relative to the actual values as measured in \citet{oliver2024}.}
    \label{CO alpha outflows}
\end{figure}

Figure \ref{CO alpha outflows} compares the four different $\alpha_{\text{CO}}$ values: $4.3\,\mathrm{M_{\odot} (\ K\ km\ s^{-1}\ pc^{2})^{-1}}$ in black, $0.8\,\mathrm{M_{\odot} (\ K\ km\ s^{-1}\ pc^{2})^{-1}}$ in green, \citet{accurso2017} in red and the conversion factor from \citet{oliver2024} in orange, which utilises the actual $H_{2}$ mass from the simulations. We find that when using the \citet{accurso2017} factor and $4.3\,\mathrm{M_{\odot} (\ K\ km\ s^{-1}\ pc^{2})^{-1}}$, the same as in Milky Way observations, the mass outflow rates are closer to the observational points, whilst if we use the actual $H_{2}$ these mass, outflow rates are lowered by 1-2 dex and this is consistent with \citet{oliver2024}, who also notes that $H_{2}$ masses within our simulations are also lower than the mass we would infer from the \citet{accurso2017} conversion factor, and this is also the case using $0.8\,\mathrm{M_{\odot} (\ K\ km\ s^{-1}\ pc^{2})^{-1}}$. \citet{stuber2021} also uses $\alpha_{\text{CO}}$ = $4.3\,\mathrm{M_{\odot} (\ K\ km\ s^{-1}\ pc^{2})^{-1}}$ and our Milky Way-mass galaxies are in good agreement here. Our simulated galaxies do not probe the starburst/ULIRG regimes which display stronger CO outflows and make up the bulk of our CO observational data, and the differences between these galaxies and our simulations is likely the cause of the discrepancy in $\dot{M}_{out}$ seen in figure \ref{mass outflow plots}.

\section{Outflow properties derived from simulation particles}
\label{outflow props from particles}
In section \ref{synthetic sepectra results} we derived outflow properties from synthetic spectra using observational methods and proceeded to test some common observational assumptions. We can also measure outflows using the simulation particles directly. However outflows within simulations are defined differently to observations and this section aims to bridge the gap between the two regimes. In simulations outflows are generally defined by a flux flowing through a given surface away from the galaxy (e.g \citealt{porter2024}) whilst observers tend to define outflows as an integrated flow of material along the line of sight. Whilst the previous section investigated outflows from an observer perspective we will now look at them from a simulator perspective. The aim of doing this is to investigate the differences between these two approaches to defining outflows and see if direct comparisons between the two are valid and what caveats may arise.

\subsection{Measuring outflows from simulation particles}
\label{pixel motivated}
We extract particles from a 3D cube with a width of $\mathrm{R_{out}}$ as derived from the corresponding images from \textsc{RADMC-3D}. In the z direction (perpendicular to the plane of the galaxy disc), a sheet of $100$\,pc thickness is placed at different distances from the disk which we vary between $0.1\,\mathrm{kpc}$ to $10\,\mathrm{H}$ where H is the scale height. Gas flowing outwards through this sheet is defined as an outflow. Gas is defined as flowing outwards if it meets the condition:

\begin{equation}
    v_{z} \times x_{z} > 0
    \label{outflow condition}
\end{equation}

where $v_{z}$ is the z-direction velocity and $x_{z}$ is the z-direction position of a particle relative to 0 at the midplane. This condition ensures that only gas flowing outwards, from either above or below the midplane, will be considered outflowing gas. From this, the mass of the outflowing gas is calculated by summing the mass of gas in a sheet which meets this condition and the outflow rate is calculated as:

\begin{equation}
    \frac{\dot{M}_{out}}{\mathrm{M_{\odot}\ yr^{-1}}} = \sum_{i} \frac{M_{i}}{\mathrm{M_{\odot}}} \frac{v_{z, i}}{\mathrm{km/s}} \frac{(\Delta z_{i}^{-1})}{\mathrm{kpc}} (3\times 10^{7}) 
    \label{MoutRate particles}
\end{equation}

where $\Delta z_{i}$ is the thickness of the sheet through which the wind flows and the $(3\times 10^{7})$ is a conversion factor from $\textrm{s}^{-1}$ to $\textrm{yr}^{-1}$. This results in a mass outflow rate, similar to those calculated in section \ref{synth spec}, with respect to the gas within the cut limits like those seen in figure \ref{m1e12 T-n cuts} for each galaxy. From this the energy and momentum outflow rates can be calculated using equations \ref{E out rate} and \ref{p out rate}. 
 
\subsection{Comparison between synthetic observations and simulation particles}
\label{rad vs pixel}
\begin{figure}
	\includegraphics[width=\columnwidth]{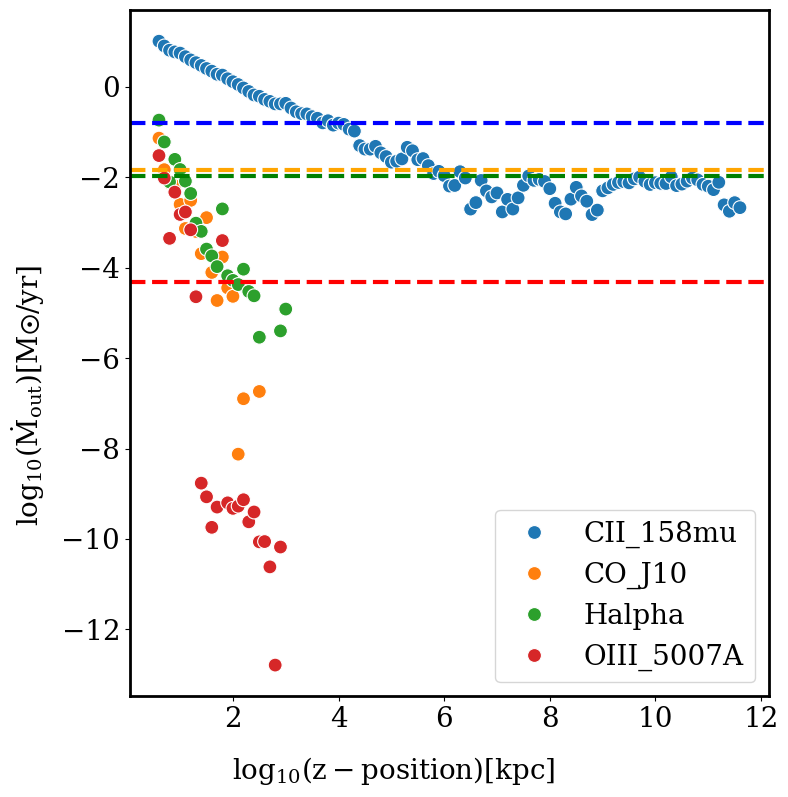}
    \caption{Mass outflow rate in a given distance bin moving from 0.1kpc to 10H in from the disk for m1e12, 100\,Myrs. Points represent outflows measured using equation \ref{MoutRate particles} whilst dashed lines represent synthetic observation mass outflow rates. [C\textsc{ii}]$_{158\rm{\mu m}}$ makes up the majority of the outflowing gas within the simulation particles, tracing gas out to a much further distance than the other phases.}
    \label{particle vs observation plots}
\end{figure}

When we measure outflow properties from spectra we are actually looking at all of the line emitting gas along the line of sight, however when we measure outflow properties using a simulation's particles (equation \ref{MoutRate particles}) we are measuring them based on a particle flux through some surface at a given height above/below the disk, meaning that what a simulator measures and observer measures are fundamentally different. Figure \ref{particle vs observation plots} shows where these two quantities align numerically, and how the mass outflow rate can vary with distance from the disk plane for [C\textsc{ii}]$_{158\rm{\mu m}}$, $\mathrm{CO}_{J(1-0)}$, $H_{\alpha}$ and [C\textsc{ii}]$_{158\rm{\mu m}}$ and compares these to the equivalent observer-like outflow rates taken for the same galaxy (m1e12,  100\,Myrs). We do this as the simulated approach and the observational approach are very different from one another, and this allows us to determine at what distance above the disk they define a mass outflow rate as equivalent.

\begin{figure*}
	\includegraphics[width=0.75\linewidth]{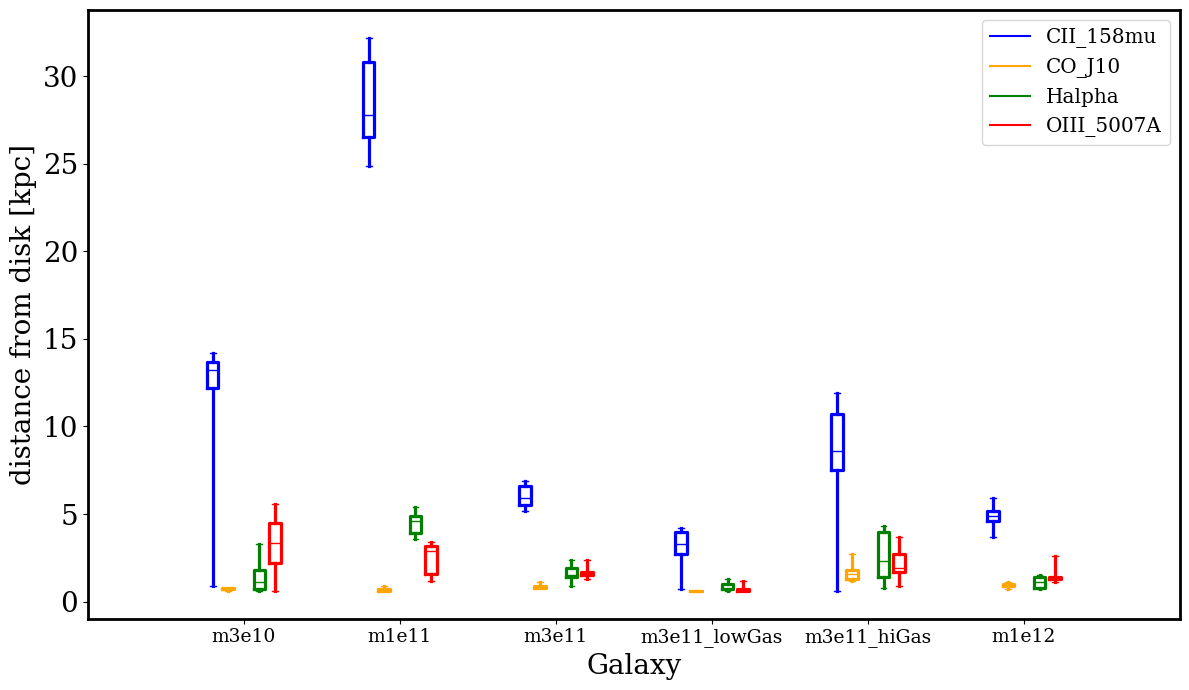}
    \caption{Boxplot showing the distance above/below the disk at which the mass outflow rates calculated from the simulation particles is equal to the mass outflow rate derived from the synthetic spectra broken down by emission line. The boxes display the median, IQ1 and IQ3 for the distance across all five of the snapshots for each galaxy. The bottom whiskers are IQ1-1.5IQR and the top whiskers are IQ3+1.5IQR. }
    \label{z-intersect_boxplot}
\end{figure*}

Figure \ref{particle vs observation plots} suggests that, in m1e12, observationally-derived outflow rates for [C\textsc{ii}]$_{158\rm{\mu m}}$ are probing gas at a distance of $\sim 5\,\mathrm{{kpc}}$, whereas the other tracers are probing gas much closer to the galaxy at $\sim 1-2\,\mathrm{kpc}$. Before we can make conclusions about what could be causing this discrepancy between the lines, we need to check whether this behaviour is systemic across all of our galaxies and all of our snapshots and we do this in figure \ref{z-intersect_boxplot}. This figure shows the distance from the disk at which the synthetic spectra outflow rates agree with the values returned by the simulations directly and this is done by comparing it to the physical measurements in figure \ref{particle vs observation plots}, for each emission line across all five snapshots for a given galaxy. We find the most drawn-out outflows are in [C\textsc{ii}]$_{158\rm{\mu m}}$ for all galaxies, reinforcing how ubiquitous this line is when compared to other emission lines looked at here.

Figure \ref{z-intersect_boxplot} shows that synthetic observations of $\dot{M}_{out}$, [C\textsc{ii}]$_{158\rm{\mu m}}$ are tracing the outflow rates at a larger distance than the other emission line tracers, and this is common across all snapshots. The behaviour of the [C\textsc{ii}]$_{158\rm{\mu m}}$ line in this case can be explained by referring to figure \ref{m1e12 T-n cuts}. [C\textsc{ii}]$_{158\rm{\mu m}}$ is present across a wider range of temperatures and densities arising in the warm neutral and cold neutral mediums. These phases are more ubiquitous within a given galaxy and therefore extend out to larger distances in comparison to other phases of gas. For the ionised tracers, an explanation may be found in \citet{gallagher2019}. They find that in $~30\%$ of galaxies hosting stellar driven outflows there is widespread star formation within the outflows themselves, as they break apart and compress together during their movement outwards. The stars formed in these outflows may then ionise the surrounding outflow gas, leading to the detection of ionised outflows further from the disk. The lack of ionised gas at distances greater than a few kpc out from the disk within our simulated galaxies implies that these stars are not being formed within our outflowing gas. The lack of AGN within our simulated galaxies may also explain the lack of long-range ionised gas as, when they are present, AGN tend to increase and dominate the excitation of outflowing gas. 

For the molecular gas there are multiple potential factors which could explain the lack of gas appearing at large heights above the disk in our simulated galaxies. The first explanation is that the molecular gas clouds may simply be falling back into the galaxy, not having been accelerated to speeds where they can move further out. Whilst molecular gas clouds are susceptible to direct acceleration via stellar radiation and cosmic rays, as the gas moves further out these methods become less effective. This is especially true in the case of molecular gas which forms dense, clumpy regions without a large surface area which allows for momentum transfer and subsequent acceleration. Using the 100\,Myrs snapshot of m1e12 as an example, in line with figure \ref{particle vs observation plots}, we have outflow velocities (see table \ref{m1e12 outflow props}) for each of our emission lines and all of these values are below the escape velocity of the 100\,Myrs snapshot of m1e12 ($v_{esc} = 250\, \mathrm{kms}^{-1}$), and the [C\textsc{ii}]$_{158\rm{\mu m}}$ gas is actually the slowest moving phase of our outflows. However this does not rule out the prospect that molecular gas is disappearing due to it falling back into the galaxy disk. Whilst the slower moving atomic gas is still being found at greater distances from the disk, this may have been moving faster than the molecular gas at the same distance, before gravity slows both tracers down and eventually pulls it back towards the disk. 

It is also likely that the molecular gas cloud is being disrupted in some way, potentially being destroyed by the hot ionised gas blowing into it. Interactions with the ionised gas can potentially drive cold molecular outflows through entrainment and in-situ formation of molecular outflows. Entrainment is the name given to cool gas being swept up and dragged along by hot, rapidly outflowing gas thanks to drag imparting a momentum on the cool cloud as the hot gas moves through \citep{veilleux2020}. This can create a multiphase outflow in line with observations. However studies of simulations investigating entrainment have found that the cold molecular clouds are likely to be rapidly destroyed when interacting with the hot wind \citep{SandR2017, SandB2015} before any molecular outflow becomes significant. Even if these clouds were to survive, it is unlikely that they would reach significant velocities due to their small cross section \citep{barragan2021}. In-situ formation is an alternative to entrainment and is the process by which cool molecular gas may be forming from the outflowing hot phase itself through cooling and condensation brought about by instabilities in the gas. \citet{richings2018agn} finds that in-situ formation of fast flowing molecular gas is possible, but their work focusses on AGN driven outflows, which are more energetic than the outflows studied here. Analysis of in-situ formation within the simulated galaxies used in this work is left for future work. 

\subsection{Discrepancies between simulation predictions and observations}
\label{disceps}
In section \ref{rad vs pixel} we saw that the molecular gas does not survive in the outflow beyond 1-2kpc (see figure \ref{particle vs observation plots}). In turn, this means there will be little to no star formation within the outflow and without stars forming in the outflowing gas, there is less ionisation occurring within the gas and therefore the mass outflow rate of the corresponding tracers will remain lower than in observed galaxies where this phenomenon is happening. [C\textsc{ii}]$_{158\rm{\mu m}}$ appears in better agreement as it extends further (figure \ref{particle vs observation plots}) out from the disk, due to it covering a larger range of temperatures and densities (figure \ref{m1e12 T-n cuts}).

\subsection{Comparison to other simulations}
\label{comparison to sims}
We can also compare our simulated galaxies to other simulations in the literature to see what differences there may be and why, alongside shedding some light on the different methods one may wish to use for analysing outflows using simulations. A good example of work which utilises a slightly different method comes from \citet{porter2024}, which studies simulations based on the FIRE subgrid models. \citet{porter2024} sets a flux surface at twice the scale height of their $z=0$ galaxies, $2H$, after testing a range of different surface heights ($H$, $2H$, $0.05R_{vir}$, $0.1R_{vir}$) and find that most quantities show a difference up to an order of magnitude between the lowest height and the largest. They find that, in line with figure \ref{particle vs observation plots}, at lower distances from the disk the flux through a given surface increases as more turbulent gas starts to be registered as an outflow. They chose $2H$ as it allows for a consistent placement of the surface between snapshots that is less disrupted by large scale outflow events and for more active galaxies at higher redshifts they choose $0.05R_{vir}$ for the same reason. They then define the mass outflow rate as:

\begin{equation}
    \frac{\dot{M}_{pix,out}}{\mathrm{M_{\odot}\ yr^{-1}}} = l_{pix}^{2} \sum_{i}\rho_{i} v_{i,z} 
    \label{porter mass outflow rate}
\end{equation}

where $\dot{M}_{pix,out}$ is the mass outflow rate using the simulation pixels, $l_{pix}$ is the side length of these pixels (750\, pc), $i$ sums over all of the pixels which overlap with the flux surface, $\rho_{i}$ is the gas density at this surface for an element $i$ and $v_{i,z}$ is the z-direction velocity (normal to the surface) for a given element $i$. They compare local outflows, which are per pixel used, and globally, which is the average value over the entire snapshot similar to how we define outflows here. In almost all of their simulations they find strong outflows driven by starburst events and find that the distribution of their mass loading factors are consistent with observations from \citet{mcquinn2019}, which find mass loading factors ranging from 0.2-7 for low mass galaxies across an SFR range of $2.5\times10^{-4}\, \mathrm{M_{\odot}yr^{-1}}$ to $4.4\, \mathrm{M_{\odot}yr^{-1}}$. \citet{porter2024} also compares to simulations from \citet{steinwandel2024} and \citet{kim+ostiker2020}, which will be covered further in this paper. They find that their results vary by up to an order of magnitude depending on the location of the flux surface and whether the outflows are defined locally or globally and that the higher the mass of a galaxy, the more energetic the outflows tend to be in order to reach a defined flux surface. They also include a circumgalactic medium, which can influence the results by providing more material for sweeping up into outflows, and provide two more simulations which utilise different methods for deriving outflow properties. 

\citet{steinwandel2024} simulates a Large Magellanic Cloud-like galaxy with halo mass $~10^{11}M_{\odot}$ with a sub-parsec resolution and a fully resolved ISM, leading a to very clear picture of a single galaxy. They measure outflow rates at different heights above the disk ($|z| = 0.5, 1, 3, 5, 10\, \mathrm{kpc}$) and they define outflowing gas the same as we do here. Instead of using the flux through a surface they calculate mass outflow rate directly at a given distance using: 

\begin{equation}
    \frac{\dot{M}_{out}}{\mathrm{M_{\odot}\ yr^{-1}}} = \sum_{i, v_{i,r}} \frac{m_{i} v_{i,r}}{dr}
    \label{steinwandel mass outflow rate}
\end{equation}

where $v_{i,r}$ is the radial velocity of an individual outflowing particle, $m_{i}$ is the gas particle mass and $dr=0.1|z|$. They measure the outflow rates using both the vertical, z based heights and radial shells based on some radius r from the centre. They find a time averaged mass outflow rate of $0.05M_{\odot}yr^{-1}$ for an average SFR of $0.05M_{\odot}yr^{-1}$ and the bulk of this mass is transported in the warm phase. This puts it above our H$\alpha$ mass outflow rates by 1 dex and above the $[\mathrm{O}\textsc{iii}]_{5007\text{\AA}}$ by up to 3 dex whilst matching up to the higher values for observations. Their work shows similar trends to our work and \citet{porter2024} and demonstrates that within simulations there is a broad range of parameters one can choose from, such as galaxy halo mass, and multiple different methods which can be used to recover mass outflow rates, and this is a good example of some work which does not use an explicit flux surface despite employing a similar method otherwise. 

Another piece of work which looks at multiple outflow phases driven by star formation is the SMAUG project \citep{kim2020}. This uses the TIGRESS framework, which use a grid-based hydrodynamics code, from \citet{kim2017} which solves ideal magneto-hydrodynamic equations to evolve the ISM over time. They define an outflow rate as a flux of the form:

\begin{equation}
    \textbf{F}_{M} = \rho v_{out}
    \label{smaug mass outflow rate}
\end{equation}

where $\textbf{F}_{M}$ is the mass flux of the outflowing gas, $\rho$ is the gas density and $v_{out}$ is the gas outflow velocity. Between them, these three methods and the one used in this paper demonstrate that there is a variety of approaches to simulating outflows and there is no standard method. One can change the definition of an outflow, the region over which you can search for outflows and the height of this region too. This is an important caveat to bear in mind for any comparisons between the results shown here and previous or future simulations.

The outflows in our simulations can be driven by a number of factors and this makes for a good comparison with \citet{yu2020}, who looks at three different starburst outflow driving mechanisms (supernovae, radiation and cosmic rays) and investigates the effect on different gas phases. They create a simulated galaxy with properties similar to M82 and derive a model which can account for all three driving mechanisms individually and all together. They find that the supernova driving mechanism leads to higher velocity, hotter outflows, which isn't what is seen in our simulated galaxies where the hot ionised phase drops off quickly. With regards to cosmic rays, they find that this driving mechanism only has a substantial effect at higher altitudes above the disk by which point most of our simulated gas has fallen back into the disk or recombined, and here the paper differs from this work as our simulations do not include cosmic ray acceleration of outflows. They also find that radiation pressure has a large influence on driving slower moving, colder outflows. However within our simulations, in the absence of AGN, supernovae are the primary driver of outflows \citep{richings2022}.

\section{Summary and Conclusions}
\label{conclusion}
To summarise, we have used a series of simulated, isolated disk galaxies based on the FIRE-2 subgrid models to compare simulated, stellar powered outflow energetics against real world observations of outflow kinematics for the cool atomic, cold molecular and hot ionised phases. The sample of galaxies used in this work represent 500\,Myr of evolution across a range of galaxy disks ranging from dwarf mass halos ($10^{10}\, \mathrm{M_{\odot}}$) to Milky Way mass halos ($10^{12}\, \mathrm{M_{\odot}}$), and their chemistry, and therefore that of the outflows, is modelling using the \textsc{chimes} non-equilibrium chemistry network. We then use \textsc{RADMC-3D} to create synthetic emission spectra to which a decomposed Gaussian is fitted, with a narrow component representing the ISM and the broad component representing outflowing gas that we use to derive outflow kinematics from this. Our results are summarised as follows:

\begin{enumerate}
    \item Comparing observationally-inferred mass outflow rates from our simulations to observations (figure \ref{mass outflow plots}) we find that we broadly reproduce the behaviour of observations, especially for [C\textsc{ii}]$_{158\rm{\mu m}}$, $[\mathrm{O}\textsc{iii}]_{5007\text{\AA}}$ and $\mathrm{H_{\alpha}}$. We find that at higher SFRs within our simulated galaxies we match observations well, with any differences confined to within $\pm1$dex depending on the observational sample.
    \item CO$_{J(1-0)}$ show differences of up to 1 dex in mass outflow rates (figure \ref{mass outflow plots}). This is likely due to differences between our simulated galaxies and observed galaxies (which are primarily starburst/ULIRGs).
    \item We test the effects of $n_{\text{e}}$ and figure \ref{SII_ratios} shows the differences between calculating $n_{\text{e}}$ using the observationally derived SII doublet ratio \citep{sanders2016} and directly measured mean $n_{\text{e}}$ values from the simulations. We find that the SII doublet ratio predicts electron densities far above the actual values in the simulations (~1.5 - 2.5 dex) in galaxies with masses $\leq 1 \times 10^{11} M_{\odot}$. Beyond this mass the electron densities appear 3 times higher when using the SII doublet as opposed to the actual values. This results in mass outflow over predictions of around 4 dex and must be considered in future observational studies.
    \item  We test the implementation of the $\alpha_{\text{CO}}$ value in section \ref{CO-H2 test} and the results of this can be seen in figure \ref{CO alpha outflows}. Comparing between a commonly used observational value of $0.8\,\mathrm{M_{\odot} (\ K\ km\ s^{-1}\ pc^{2})^{-1}}$, values derived from \citet{accurso2017} and measurements from \citet{oliver2024} (which uses the same simulated galaxies), we find that our outflows are sensitive to $\alpha_{\text{CO}}$, especially as $M_{200}$ increases. We find that the largest differences appear in m1e10, with the \citet{accurso2017} and $0.8\,\mathrm{M_{\odot} (\ K\ km\ s^{-1}\ pc^{2})^{-1}}$ $\alpha_{\text{CO}}$ values differing by ~3 dex, whilst in m1e12 the differences are ~2 dex between \citet{oliver2024} and \citet{accurso2017}.
    \item In section \ref{CII temp test} we test the influence of temperature on our [C\textsc{ii}]$_{158\rm{\mu m}}$ mass outflow rates. We begin by using the median value of the temperature range described in \citet{romano2023}, which is 130\,K. We compare this to the mass outflow calculated using the temperature of the [C\textsc{ii}]$_{158\rm{\mu m}}$ gas derived directly from the simulations. We find that the equation used for calculating outflowing mass is not sensitive to the temperatures of the gas within our simulations, with little difference between the 130\,K outflows and those using the actual temperatures.
    \item To further understand our observationally-derived outflow properties we compare our simulated observations to actual simulation values, by comparing the outflow properties across the different phases as measured using the simulation particles themselves. We find that there is a rapid drop off in the molecular and ionised phases, whilst the [C\textsc{ii}]$_{158\rm{\mu m}}$ gas dominates the $\dot{M}_{out}$ at all distances from the disk (figure \ref{particle vs observation plots}). For the ionised phase, the drop off may be explained by a lack of star formation within the outflows themselves, a process which is explained in \citet{gallagher2019}. Freshly formed stars within the outflow would subsequently ionise the outflowing atomic gas and increase the distribution of ionised outflows.
\end{enumerate}

In the future, this work can be extended to larger cosmological simulations. This would allow us to investigate outflows over a far larger number of galaxies as well as considering the influence of the circumgalactic medium and gaseous inflows on the energetics investigated here. Having access to a larger scale would also allow a more in-depth study of outflows' lifecycles alongside being able to track their ultimate fate.

\section*{Acknowledgements}

We thank the reviewer, Martin Rey, for his constructive feedback which improved this paper. CAFG was supported by NSF through grants AST-2108230 and AST-2307327; by NASA through grants 21-ATP21-0036 and 23-ATP23-0008; and by STScI through grant JWST-AR-03252.001-A. D.A.A. acknowledges support from NSF grant AST-2108944 and CAREER award AST-2442788, NASA grant ATP23-0156, STScI grants JWST-GO-01712.009-A, JWST-AR-04357.001-A, and JWST-AR-05366.005-A, an Alfred P. Sloan Research Fellowship, and Cottrell Scholar Award CS-CSA-2023-028 by the Research Corporation for Science Advancement. This work used the DiRAC@Durham facility managed by the Institute for Computational Cosmology on behalf of the STFC DiRAC HPC Facility (www.dirac.ac.uk). The equipment was funded by BEIS capital funding via STFC capital grants ST/K00042X/1, ST/P002293/1, ST/R002371/1, and ST/S002502/1, Durham University and STFC operations grant ST/R000832/1. DiRAC is part of the National eInfrastructure. This work also used Viper, the University of Hull High Performance Computing Facility. 

\section*{Data Availability}
The data underlying this article will be shared on reasonable request to the corresponding authors. 

A public version of the CHIMES astrochemistry code can be found at https://richings.bitbucket.io/chimes/home.html, and a public version of the GIZMO hydrodynamics code can be found at http://www.tapir.caltech.edu/~phopkins/Site/GIZMO.html. 




\bibliographystyle{mnras}
\bibliography{bibliography} 




\appendix


\bsp	
\label{lastpage}
\end{document}